\documentclass[]{achemso}

\usepackage[version=3]{mhchem} % Formula subscripts using \ce{}
\usepackage{bm}
\usepackage{siunitx}

\makeatletter
\DeclareTextCompositeCommand{\r}{OT1}{A}{%
  \leavevmode\vbox{%
    \offinterlineskip
    \ialign{\hfil##\hfil\cr\char23\cr\noalign{\kern-1.15ex}A\cr}%
  }%
}
\makeatother

\newcommand{\mycomment}[1]{}

\author{Ikuma Kohata}
\affiliation[Universty of Tokyo]{Department of Mechanical Engineering, The University of Tokyo, 7-3-1 Hongo, Bunkyo-ku, Tokyo 113-8656, Japan}

\author{Ryo Yoshikawa}
\affiliation[Universty of Tokyo]{Department of Mechanical Engineering, The University of Tokyo, 7-3-1 Hongo, Bunkyo-ku, Tokyo 113-8656, Japan}

\author{Kaoru Hisama}
\affiliation[Shinshu University]
{Research Initiative for Supra-Materials, Shinshu University, 4-17-1 Wakasato, Nagano, Nagano 380-8553, Japan}

\author{Christophe Bichara}
\affiliation[Marseille]
{Aix Marseille Univ, CNRS, Centre Interdisciplinaire de Nanoscience de Marseille, Marseille, France}

\author{Keigo Otsuka}
\affiliation[University of Tokyo]
{Department of Mechanical Engineering, The University of Tokyo, 7-3-1 Hongo, Bunkyo-ku, Tokyo 113-8656, Japan}

\author{Shigeo Maruyama}
\affiliation[University of Tokyo]
{Department of Mechanical Engineering, The University of Tokyo, 7-3-1 Hongo, Bunkyo-ku, Tokyo 113-8656, Japan}
\altaffiliation
{School of Mechanical Engineering, Zhejiang University, Hangzhou 310027, People’s Republic of China}

\email{maruyama@photon.t.u-tokyo.ac.jp}

\title[An \textsf{achemso} demo]
  {Edge Dynamics in Iron-Cluster Catalyzed Growth of Single-Walled Carbon Nanotubes Revealed by Molecular Dynamics Simulations based on a Neural Network Potential}


\begin{document}

%%%%%%%%%%%%%%%%%%%%%%%%%%%%%%%%%%%%%%%%%%%%%%%%%%%%%%%%%%%%%%%%%%%%%
%% The "tocentry" environment can be used to create an entry for the
%% graphical table of contents. It is given here as some journals
%% require that it is printed as part of the abstract page. It will
%% be automatically moved as appropriate.
%%%%%%%%%%%%%%%%%%%%%%%%%%%%%%%%%%%%%%%%%%%%%%%%%%%%%%%%%%%%%%%%%%%%%

\begin{tocentry}

\includegraphics[width=\textwidth]{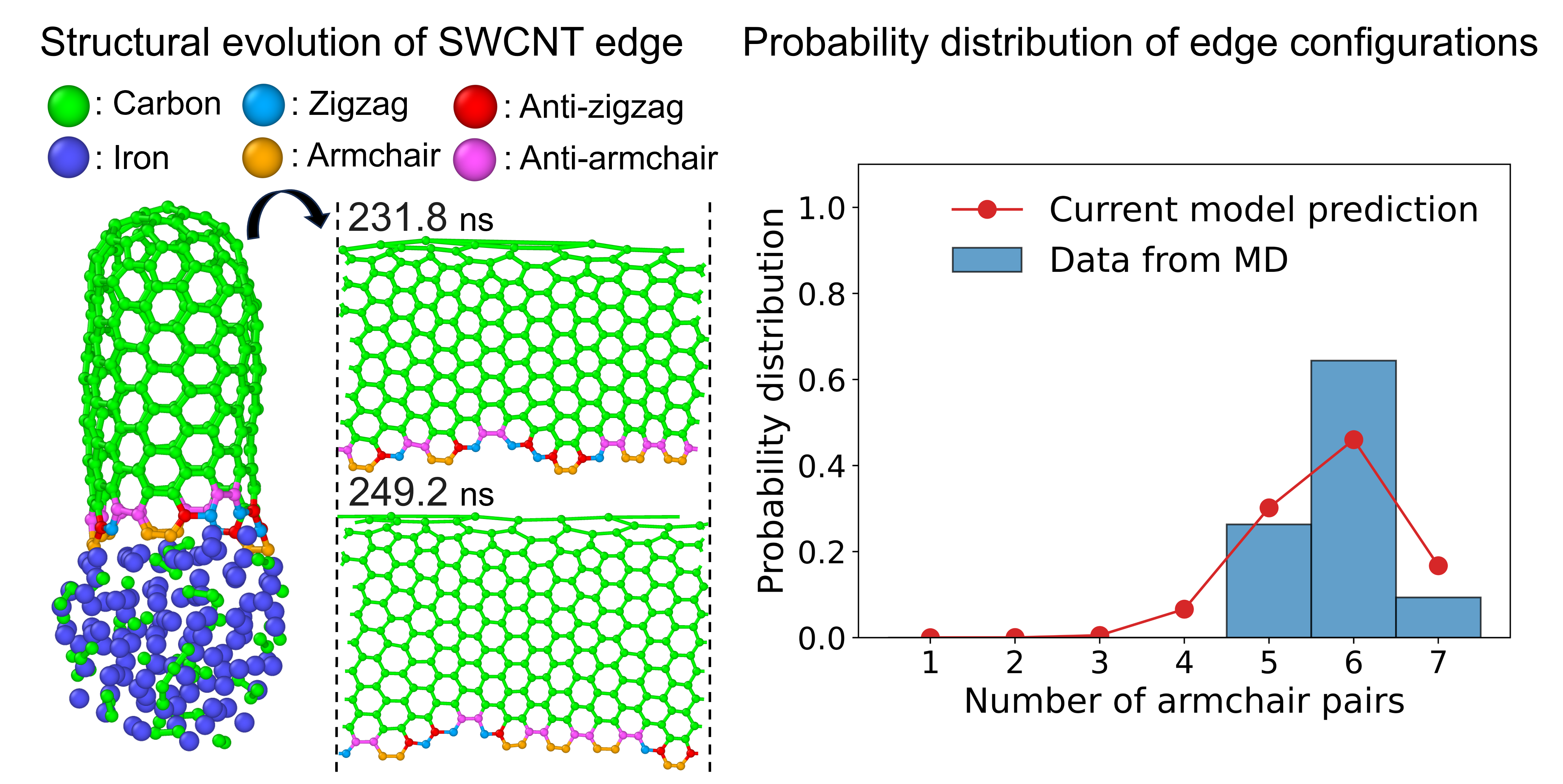}

\end{tocentry}

%%%%%%%%%%%%%%%%%%%%%%%%%%%%%%%%%%%%%%%%%%%%%%%%%%%%%%%%%%%%%%%%%%%%%
%% The abstract environment will automatically gobble the contents
%% if an abstract is not used by the target journal.
%%%%%%%%%%%%%%%%%%%%%%%%%%%%%%%%%%%%%%%%%%%%%%%%%%%%%%%%%%%%%%%%%%%%%
\begin{abstract}
Given the high potential for applications utilizing the unique properties of single-walled carbon nanotubes (SWCNTs), there is considerable enthusiasm for addressing the challenges associated with synthesizing SWCNTs with specific chirality. To elucidate the mechanisms that determine the chirality of SWCNTs during growth, intensive efforts have been devoted to classical molecular dynamics (MD) simulations. However, the mechanism of chirality determination has not been fully clarified, which can partly be attributed to the limited accuracy of empirical interatomic potentials in reproducing the behavior of carbon and metal atoms. In this work, we develop a neural network potential (NNP) for carbon-metal system to accurately describe the SWCNT growth, and perform MD simulations of SWCNT growth using the NNP. The MD simulations illustrate the defect-free, chirality-definable growth of SWCNTs, highlighting the dynamic rearrangement of edge configurations and the consistency between the probability of edge configuration appearance and the entropy-driven edge stability model proposed here. It is also shown that the edge defect formation is induced by vacancy and suppressed by vacancy healing through adatom diffusion on the SWCNT edges. These results provide insights into the edge formation thermodynamics and kinetics of SWCNTs, an important clue to the chirality-controlled synthesis of SWCNTs.
\end{abstract}

%%%%%%%%%%%%%%%%%%%%%%%%%%%%%%%%%%%%%%%%%%%%%%%%%%%%%%%%%%%%%%%%%%%%%
%% Start the main part of the manuscript here.
%%%%%%%%%%%%%%%%%%%%%%%%%%%%%%%%%%%%%%%%%%%%%%%%%%%%%%%%%%%%%%%%%%%%%
\section{Introduction}
Single-walled carbon nanotubes (SWCNTs) have great potential \cite{doi:10.1021/acs.chemrev.0c00395,doi:10.1021/acs.chemrev.9b00835} as a next-generation material for nanoscale devices such as sensors, optoelectronic devices, and high-performance logic transistors\cite{doi:10.1126/science.abp8278,doi:10.1126/science.aba5980,Hills2019} because of their excellent physical properties\cite{doi:10.1126/science.aba5980,doi:10.1126/science.1222453}. The electronic structure of SWCNTs dramatically varies from metallic to semiconducting depending on their chirality\cite{doi:10.1063/1.107080,Wilder1998,Odom1998}, and chirality-controlled synthesis is important to obtain chirality-pure SWCNT samples without contamination and damage caused by sorting, and to realize high-end SWCNT-based devices.
Many experimental studies have been carried out to achieve chirality-controlled growth of SWCNTs\cite{doi:10.1021/acs.chemrev.9b00835}. Although these impressive results shed light on chirality control, further understanding of the chirality-determining SWCNT growth mechanism is needed to produce higher purity SWCNT samples with arbitrary chirality in large quantity.

In the past few decades, theoretical models have been proposed to account for the chirality distribution of SWCNTs\cite{doi:10.1073/pnas.0811946106,doi:10.1126/science.aat6228}. These models successfully relate the state of the interface between the SWCNT edge and the catalyst to the chirality distribution, but do not provide a detailed atomistic mechanism in the edge formation process because they are based on some assumptions about the process of carbon addition to the SWCNT edges. To reveal such mechanism, many molecular dynamics (MD) simulations have also been performed. In particular, classical MD simulations, which are suitable for observing nanotube growth on a long time scale, have demonstrated a series of growth processes from the nucleation to the sidewall elongation\cite{SHIBUTA2003381,doi:10.1063/1.1770424,doi:10.1021/jp046645t,Zhao_2005,DING2004309,Ding_2006,doi:10.1021/nl072431m}. 
As a first approach to the discussion of chirality in MD simulation, Neyts et al.\cite{doi:10.1021/nn102095y,doi:10.1021/ja204023c} performed hybrid atomistic simulations of MD and force-biased Monte Carlo based on the reactive force field (ReaxFF). They defined the local chirality by the diameter and the angle of the hexagon to the axial direction of SWCNTs. In their simulations of the initial growth process of SWCNTs, chirality changes from zigzag to armchair were observed.
Yoshikawa et al.\cite{doi:10.1021/acsnano.8b09754} carried out MD simulations with a lower carbon supply rate, and assigned the chirality to SWCNTs by defect-free growth using the Tersoff-type bond-order potential. This chirality-definable MD simulation demonstrated the kink-running process at SWCNT edges predicted by the dislocation theory of SWCNT growth \cite{doi:10.1073/pnas.0811946106}. However, the chiralities of the observed SWCNTs were only zigzag and near-zigzag, which is contrary to the experimental reports of near-armchair preferential growth\cite{doi:10.1021/ja070808k,C2NR32276E,He2013,doi:10.1021/ja106937y}. This suggests that the bond-order potential shows an excessive preference for zigzag edges, as reported in a theoretical study\cite{PhysRevLett.105.235502}, and that the preference for armchair edges over zigzag edges predicted by the DFT calculation is not correctly reproduced by the bond-order potential.

Currently, there are two reasons why it is difficult to accurately simulate SWCNT growth. Firstly, the time scale of MD simulations is smaller than that of experiments, and there is not enough time for defects to heal during simulated SWCNT growth. Secondly, the energetics of SWCNTs including defects is not accurately described by classical interatomic potentials. For more accurate description of atomistic behavior in SWCNT growth, it is necessary to perform long MD simulations using an accurate interatomic potential.

Recently, the application of artificial neural networks to interatomic potentials has been actively studied to overcome the limitations of conventional interatomic potentials and such interatomic potentials are called neural network potentials (NNPs). Since the idea of symmetry functions to represent high-dimensional potential energy surfaces (PESs) with translational and rotational invariance was proposed by Behler and Parrinello\cite{PhysRevLett.98.146401}, a number of approaches have been developed to construct NNPs in the field of materials science\cite{MISHIN2021116980}. Among the NNPs, the deep potential\cite{PhysRevLett.120.143001} is one of the most popular methods because of its high accuracy and high computational efficiency. Hedman et al. have performed MD simulations of SWCNT growth using the deep potential at the same time, with a different training dataset, independently of this work \cite{Hedman2024}, and healing of interface defects was extensively studied.

In this study, we develop an NNP based on the deep potential for C-Fe system to accurately simulate the SWCNT growth. 
The NNP is trained against iteratively collected 44000 structures, achieving good accuracy in the defect formation energy of graphene and the thermodynamic stability of C-Fe nanoparticles. 
Using the developed NNP, we perform MD simulations of the defect-free SWCNT growth by supplying carbon atoms to Fe nanoparticles. 
The MD simulations demonstrate the process of defect-free chirality-definable SWCNT growth with dynamic rearrangement of edge configurations, and the consistency between the probability of edge configuration appearance and the entropy-driven edge stability predicted from the nanotube-catalyst interfacial energy.
We also discuss the mechanism of edge defect formation and healing, which leads to the defect-free growth.

\section{Results and discussion}

\subsection{Validation of NNP for Fe-catalyzed SWCNT Growth}
\hspace{\parindent}
The growth process of SWCNTs consists of the dynamic formation of carbon rings on the catalyst surface through recombining carbon-carbon bonds. In order to reproduce the SWCNT growth by MD simulation, interatomic potentials need to predict properties of various morphology of C-Fe system including unstable structures. In this study, we used classical MD simulation for sampling dataset structures and trained an NNP that can accurately predict the energies and forces of various structures including molecules, chains, clusters, graphene, and nanotubes. Before focusing on the mechanism of SWCNT growth, we validated the NNP from the perspectives of defect formation energy and edge free energy of graphene, thermodynamic behavior of C-Fe nanoparticles. 

%\subsection{Energetics of graphene}
\subsubsection{Defect formation energies of graphene}
\hspace{\parindent}To simulate SWCNT growth, a key parameter is the stability of defects.
To validate the NNP, we checked the predictive performance of defect formation energies of graphene for the defective graphene dataset (Figure S1b), and compared it with conventional interatomic potentials. Defect formation energy $D$ was defined as 
\begin{equation}
D = E_{d}/n_{d}-E_{g}/n_{g}.
\end{equation}
Here, $E_{d}$ and $E_{g}$ are the total energy of defective and ideal graphene structure, respectively. $n_{d}$ and $n_{g}$ are the number of atoms in the supercell of the defective and ideal graphene, respectively. The comparison of defect formation energies between density functional theory (DFT) calculation and the NNP is plotted in Figure \ref{fig:edgeenergy}a. The same plots for Tersoff-type potential\cite{doi:10.1021/acs.jpcc.7b12687} and ReaxFF\cite{doi:10.1021/jp004368u,doi:10.1021/jp9035056}, employed in previous SWCNT growth simulations, are also shown. The plots indicate that the developed NNP much more accurately predicted the defect formation energy than the other two interatomic potentials. The Tersoff-type potential underestimated the defect formation energy, which is considered to result in more frequent defect formation during SWCNT growth simulations\cite{doi:10.1021/acs.jpcc.7b12687,doi:10.1021/acsnano.8b09754}. In contrast, the ReaxFF potential overestimated the defect formation energy and accordingly the stability of graphene, which may have prevented cap lift-off driven by pentagon formation during the initial stages of SWCNT growth.

\subsubsection{Edge energy of graphene}
\begin{figure}[H]
\begin{minipage}[c]{0.4\columnwidth}
\centering
\includegraphics[clip,scale=0.45]{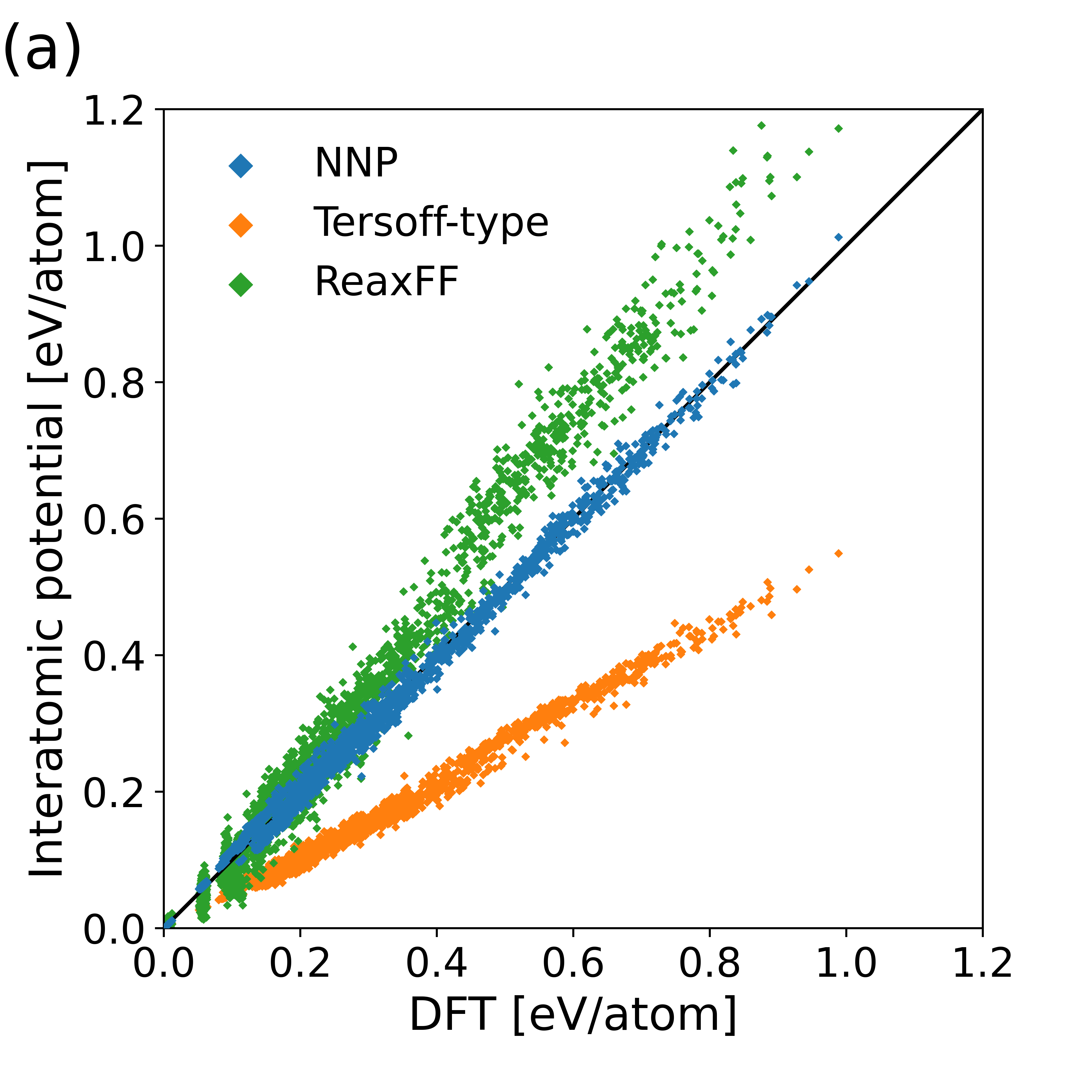}
\end{minipage}
\begin{minipage}[c]{0.4\columnwidth}
\centering
\includegraphics[clip,scale=0.45]{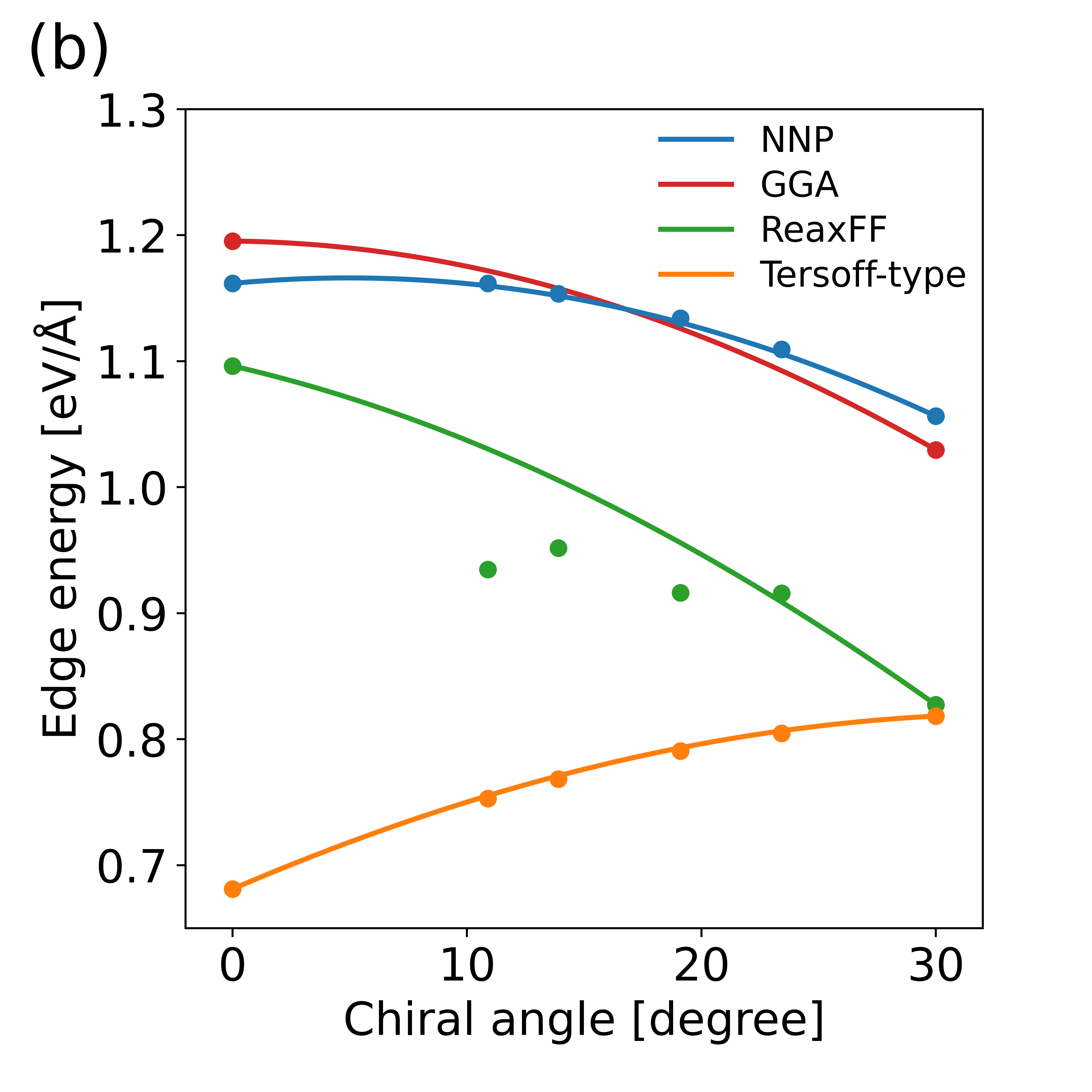}
\end{minipage}
\caption{ Comparison of (a) defect formation energy of graphene and (b) edge energy of graphene as a function of chiral angle between DFT with GGA, the developed NNP, Tersoff-type potential\cite{doi:10.1021/acs.jpcc.7b12687}, and ReaxFF\cite{doi:10.1021/jp004368u,doi:10.1021/jp9035056}. The edge energies of graphene were directly computed (dots) and fitted with Eq. \ref{eq:edge} (lines).}
\label{fig:edgeenergy}
\end{figure}

\hspace{\parindent}The edge energy is considered to be an important factor of SWCNT chirality selectivity\cite{PhysRevLett.105.235502,doi:10.1073/pnas.1207519109}. To verify the NNP and discuss the chirality selectivity in terms of thermodynamic stability, we checked the chiral angle dependence of the edge energy of graphene.
To calculate the edge energy by the NNP, we modeled graphene nanoribbons (GNRs) in the cell with one periodic direction and a vacuum space of 15 \r{A} in the other two non-periodic directions. We relaxed them with the conjugate gradient method.
The edge energy $\gamma$ is defined as
\begin{equation}
\gamma=(E_{total}-N_{C}E_{C}-N_{e}\mu)/2L,
\end{equation}
where $E_{total}$ is the total energy of the supercell, $N_{C}$ is the total number of carbon atoms in the supercell, $E_{C}$ is the energy of carbon atoms in the graphene lattice, $N_{e}$ is the number of terminating atoms, $\mu$ is the chemical potential of the terminating element, and $L$ is the periodic length of graphene nanoribbon as defined in a previous theoretical study\cite{PhysRevLett.105.235502}. 
We computed the edge energy of the GNRs plotted as a function of chiral angle, as shown in Figure \ref{fig:edgeenergy}b. For comparison, we also plotted the analytical curve \cite{PhysRevLett.105.235502} expressed as 
\begin{equation}
\gamma^{\prime}(\chi)=2\gamma^{\prime}_{A}\sin{(\chi)}+2\gamma^{\prime}_{Z}\sin{(30^{\circ}-\chi)},
\label{eq:edge}
\end{equation}
where $\chi$ is the chiral angle, $\gamma^{\prime}_{A}$ and $\gamma^{\prime}_{Z}$ are the edge energies of armchair and zigzag edge, respectively. On pristine graphene edges, the energies computed with the NNP showed higher stability of large chiral angles, which agreed with the generalized gradient approximation (GGA) of DFT. On the other hand, those computed with the Tersoff-type potential indicated higher stability of small chiral angles, which was considered to result in the zigzag preferential growth by previous MD simulations\cite{doi:10.1021/acsnano.8b09754}.

\subsubsection{Thermodynamic properties of Fe}
\hspace{\parindent}The influence of the catalyst phase on SWCNT growth has been reported in experimental studies\cite{doi:10.1021/nn304064u,doi:10.1021/acsnano.0c05542}. In order to accurately simulate SWCNT growth, the thermodynamic behavior of catalyst nanoparticles must be well considered. 
This section discusses the thermodynamic behavior of Fe bulk crystals and nanoparticles.
First, we estimated the melting point of bulk body centered cubic (bcc) Fe by two-phase simulation\cite{PhysRevB.49.3109} using the NNP. For two-phase simulation, we prepared the initial structure of bulk Fe with a bcc-liquid interface by melting the upper half of a bcc supercell of $28.52\times28.52\times74.17$ \r{A} with 5200 atoms at 5000 K (Figure S3a). Starting with the initial structures, we carried out MD simulations for 200 ps under NPT ensemble at 1 atm with different temperatures. The final structures at 1850, 1887.5, and 1900 K are shown in Figure S3b-d, and the potential energy was well converged at each temperature, as shown in Figure S3e. The equilibrium temperature was 1887.5$\pm$12.5 K, which agrees well with the experimental melting point of 1811 K\cite{Kittel}. In addition, the latent heat of melting was estimated as 0.15 eV/atom from the difference of the averaged potential energies between solid and liquid, which also agrees with the experimental value 0.143 eV/atom\cite{doi:10.1063/1.1699435} (Figure S3f).

\begin{figure}[H]
\begin{minipage}[b]{1.0\columnwidth}
\centering
\includegraphics[clip,scale=0.45]{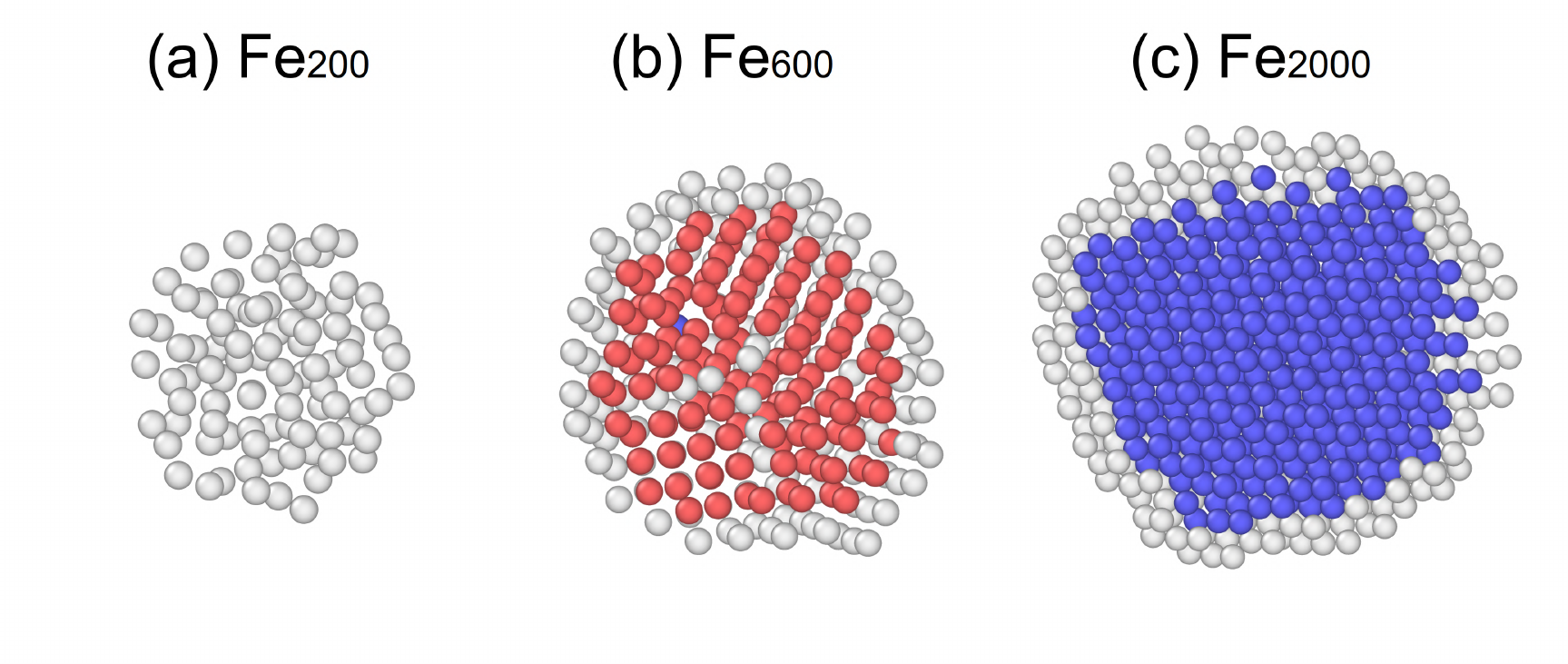}
\end{minipage}\\
\begin{minipage}[b]{0.40\columnwidth}
\centering
\includegraphics[clip,scale=0.45]{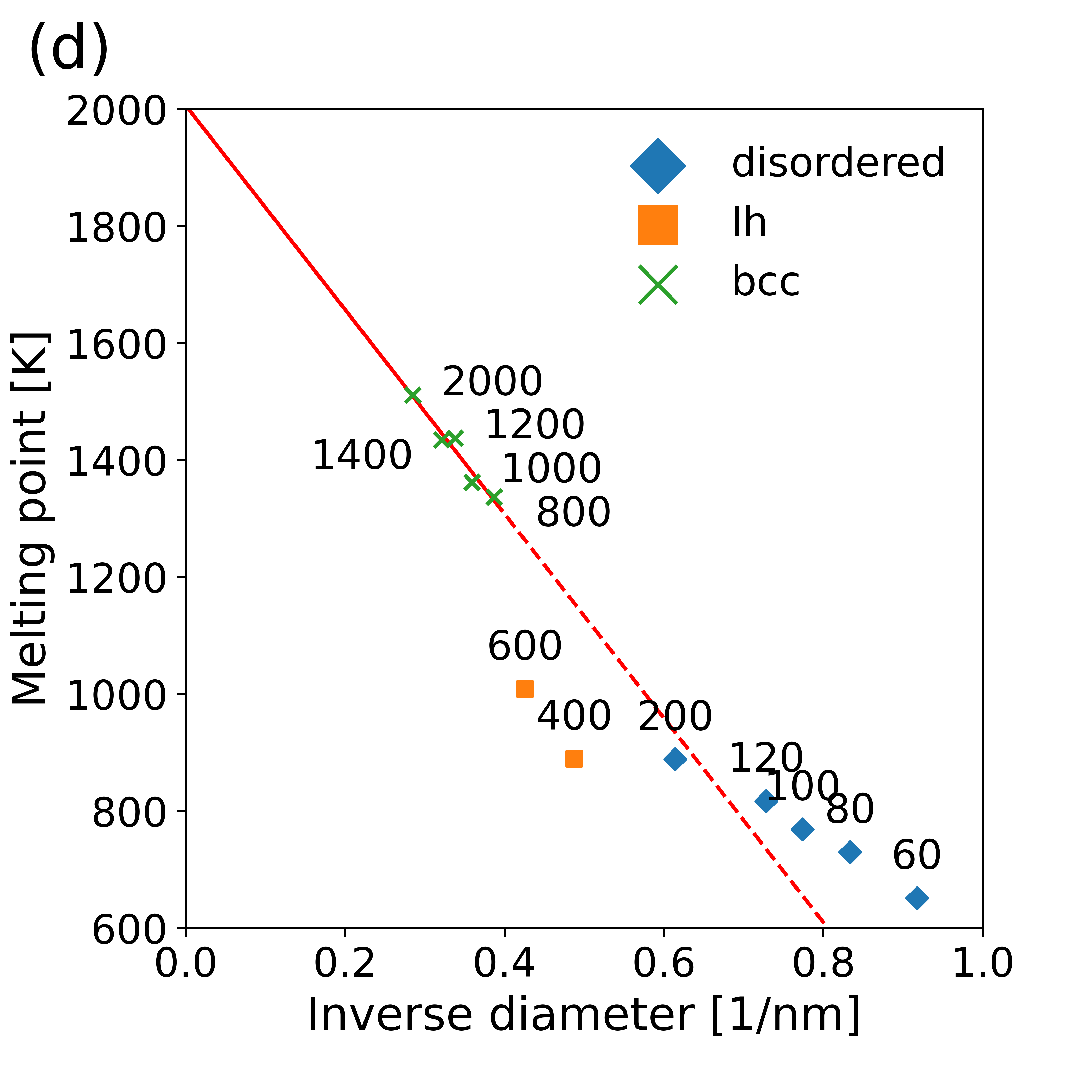}
\end{minipage}
\begin{minipage}[b]{0.40\columnwidth}
\centering
\includegraphics[clip,scale=0.45]{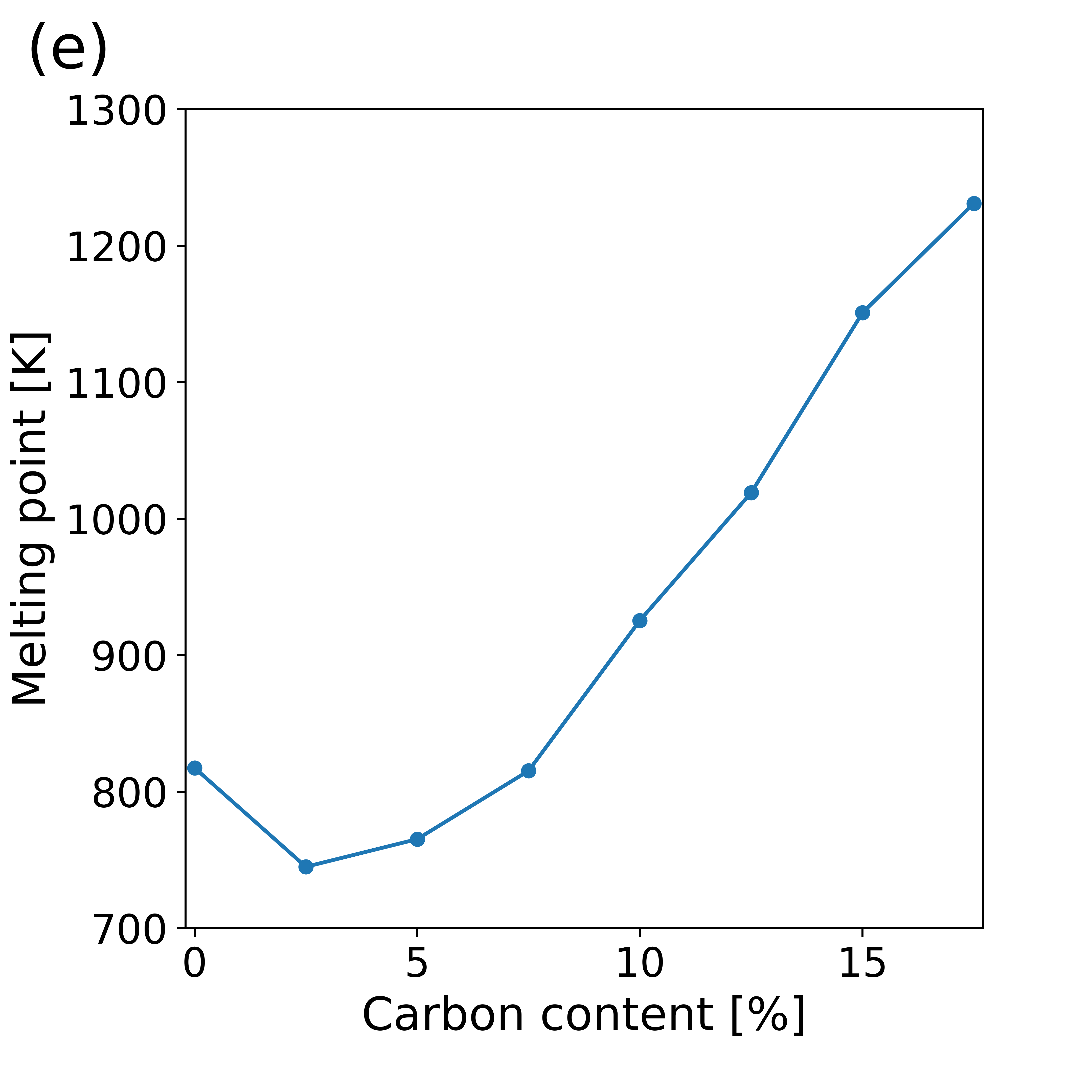}
\end{minipage}
\caption{Thermodynamic behaviors of Fe nanoparticles. (a-c) Snapshots of cross sectional views of annealed structures of Fe$_{200}$, Fe$_{600}$ and Fe$_{2000}$ nanoparticles. The blue and red balls denote atoms in bcc and Ih structure, respectively. The white balls denote atoms out of bcc and Ih structures. The structures were categorized by the common neighbor analysis\cite{doi:10.1021/j100303a014}. (d) Melting point of Fe nanoparticles as a function of inverse diameter. The red line is linear fits to data points for bcc Fe nanoparticles. (e) Melting points of Fe$_{120}$C$_{n}$ nanoparticles obtained by adding up to 17.5 \% of carbon atoms. }
\label{fig:linde}
\end{figure}

Next, we verified the thermodynamic behavior of Fe nanoparticles by monitoring the Lindemann index representing the root-mean-square relative bond-length fluctuation
\begin{equation}
  \delta=\frac{2}{N(N-1)}\sum_{i<j}\frac{\sqrt{\langle{r_{ij}^2}\rangle-\langle{r_{ij}}\rangle^2}}{\langle{r_{ij}}\rangle},
\end{equation}
where $r_{ij}$ is the distance between atom $i$ and $j$, N is the number of particles and the bracket is the time average calculated over MD simulation at constant temperature $T$. The melting point was determined as the temperature at which the Lindemann index changed discontinuously. Previous studies used a threshold value in the range of 0.08-0.15 as the melting point criteria\cite{doi:10.1116/1.1752895,doi:10.1063/1.2991435,RADEV2002242,DUAN200757,SHIBUTA2007265,PhysRevB.77.115450}. However, in the case of small nanoparticles, the solid and liquid phases coexist over a wide temperature range, and the criteria for the melting point have not been well established. 
In this study, to estimate the melting point, a hyperbolic function was fitted to the data point of the Lindemann index. The mathematical form of the hyperbolic function is expressed as
\begin{equation}
  f(t)=\frac{(at+b)\exp{(\frac{t-v}{s})}+
  (ct+d)\exp{(-\frac{t-v}{s})}}{\exp{(\frac{t-v}{s})+\exp{(-\frac{t-v}{s})}}}+w,
\end{equation}
where $a$, $b$, $c$, $d$, $e$, $s$, $v$, and $w$ are fitting parameters. The melting point was determined as $T_{m}=w$ where the inflection point of the fitted hyperbolic curve is located (see Figure S4, Supporting Information). For MD simulation to obtain the Lindemann index, the initial structures were prepared by melting Fe nanoparticles at 2000 K and then annealing them to 0 K at 1 K/ps using the NNP. The annealed structures of Fe$_{200}$, Fe$_{600}$ and Fe$_{2000}$ are shown in Figures \ref{fig:linde}a-c. The stable phase of Fe nanoparticles changed depending on the size. For large nanoparticles ($N>$700), the bcc structure was the stable phase as well as bulk Fe. However, for medium-size nanoparticles (300$<N<$700), icosahedral (Ih) was found stable. No regular structures were not observed for smaller nanoparticles ($N<$300) as not all possible magic number structures were used as a starting point in this first approach. This phase transition from bcc to Ih by size, which was considered to be due to surface effect, is consistent with high-resolution transmission electron microscopy observation of Ih Fe nanoparticles\cite{doi:10.1021/nl8037294}. Starting from the annealed structures, MD simulations were performed for 2 ns at constant temperature $T$ and the Lindemann index was then obtained as the average over the latter half 1 ns of MD simulations. According to the Gibbs-Thomson equation, the decrease in the melting point is inversely proportional to the diameter. As shown in Figure \ref{fig:linde}d, the melting point followed the equation for large nanoparticles of bcc Fe. However, the melting point discontinuously changed by 300 K between Fe$_{600}$ and Fe$_{800}$. This discontinuity of the melting point corresponded to the phase transition from bcc to Ih. 

In addition, the melting points of Fe$_{120}$C$_{n}$ were calculated, as shown in Figure \ref{fig:linde}e. The melting point decreased from 814 K to 740 K with $\sim$2.5 \% carbon content, while it increased with more carbon content. The observed V-shape diagram is consistent with the phase diagram of bulk Fe-C\cite{2phase} and previous computational studies\cite{doi:10.1116/1.1752895,PhysRevB.75.205426}. The following simulations of SWCNT growth are performed under conditions where the Fe catalysts are fully liquid.

\subsection{SWCNT Growth Simulation}
\subsubsection{Defect free growth of SWCNTs}
\begin{figure}[H]
\begin{minipage}[b]{1.0\columnwidth}
\centering
\includegraphics[clip,scale=0.48]{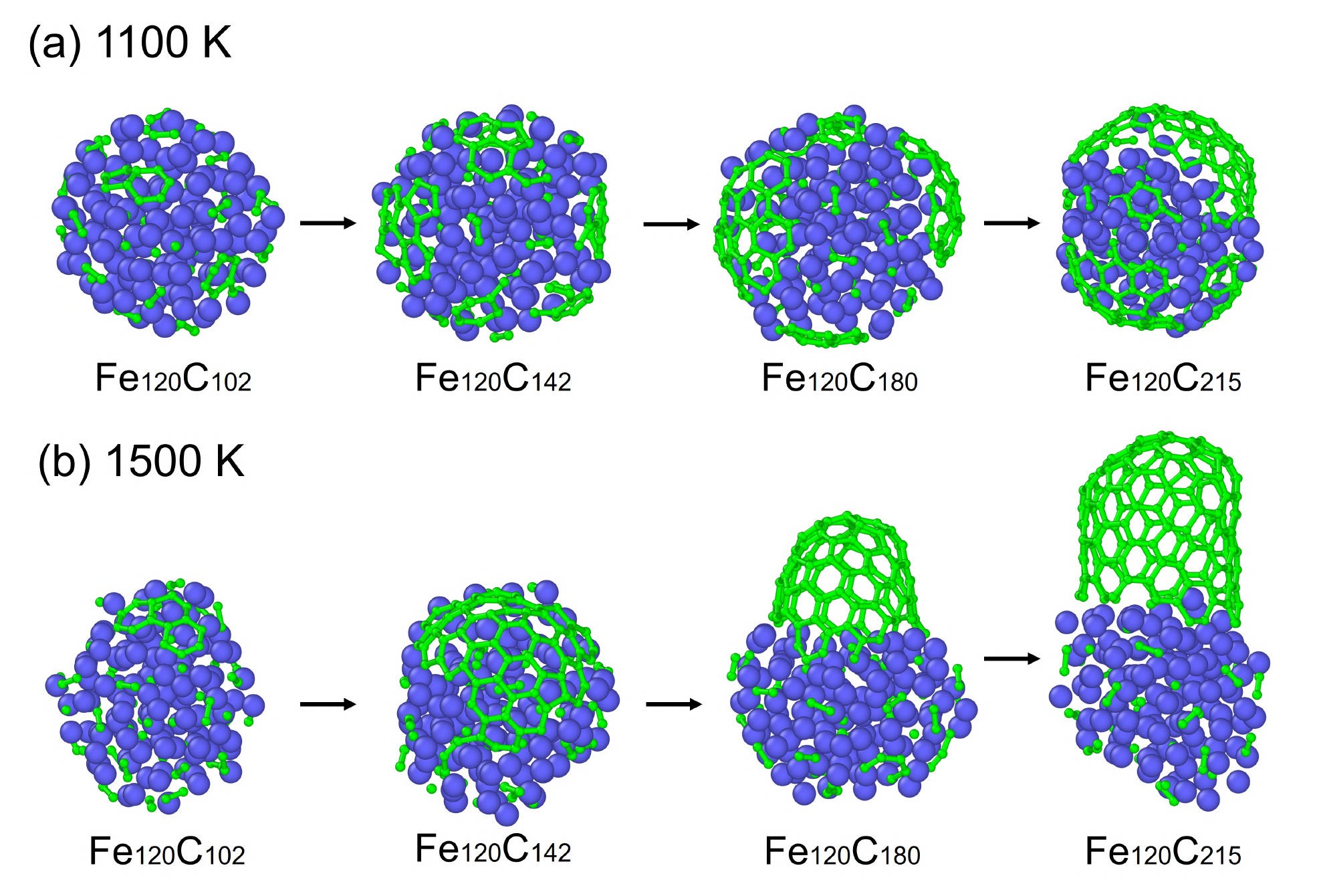}
\end{minipage}
\caption{Snapshots of the initial stages of SWCNT growth at (a) 1100 K and (b) 1500 K. The blue and green atoms denote iron and carbon, respectively.} 
\label{fig:trend}
\end{figure}

\hspace{\parindent}Using the developed NNP, we carried out SWCNT growth simulations by supplying C atoms to Fe nanoparticles in the same way as the previous SWCNT growth simulations\cite{doi:10.1021/acs.jpcc.7b12687,doi:10.1021/acsnano.8b09754}.
We firstly performed twenty MD simulations of SWCNT growth on Fe$_{55}$, Fe$_{80}$, Fe$_{100}$ and Fe$_{120}$ nanoparticles at $T$=$1100$-$1500$ K in 100 K increments to investigate the growth trends by temperature (Figure S5). Figure \ref{fig:trend} shows the typical growth on Fe$_{120}$ nanoparticles at (a) 1100 K, and (b) 1500 K. At lower temperatures (1100-1300 K), the mobility of carbon atoms was limited, preventing their incorporation into a single large graphitic island. In these conditions, isolated hexagons nucleated simultaneously at several sites on the surface. Multiple graphitic islands then formed from these hexagons, covering the surface of the nanoparticles without lift-off and leading to the deactivation.
Conversely, at high temperatures (1400-1500 K), a single hexagon was formed, enabling the sequential addition of carbon atoms to create a unified large island and the cap formation. 
Note that the threshold temperature is supposed to depend on the carbon supply rate, and may differ from the experimental value because the supply rate in this simulation is significantly higher than experiments.

\begin{figure}[H]
\begin{minipage}[b]{1.0\columnwidth}
\centering
\includegraphics[clip,scale=0.5]{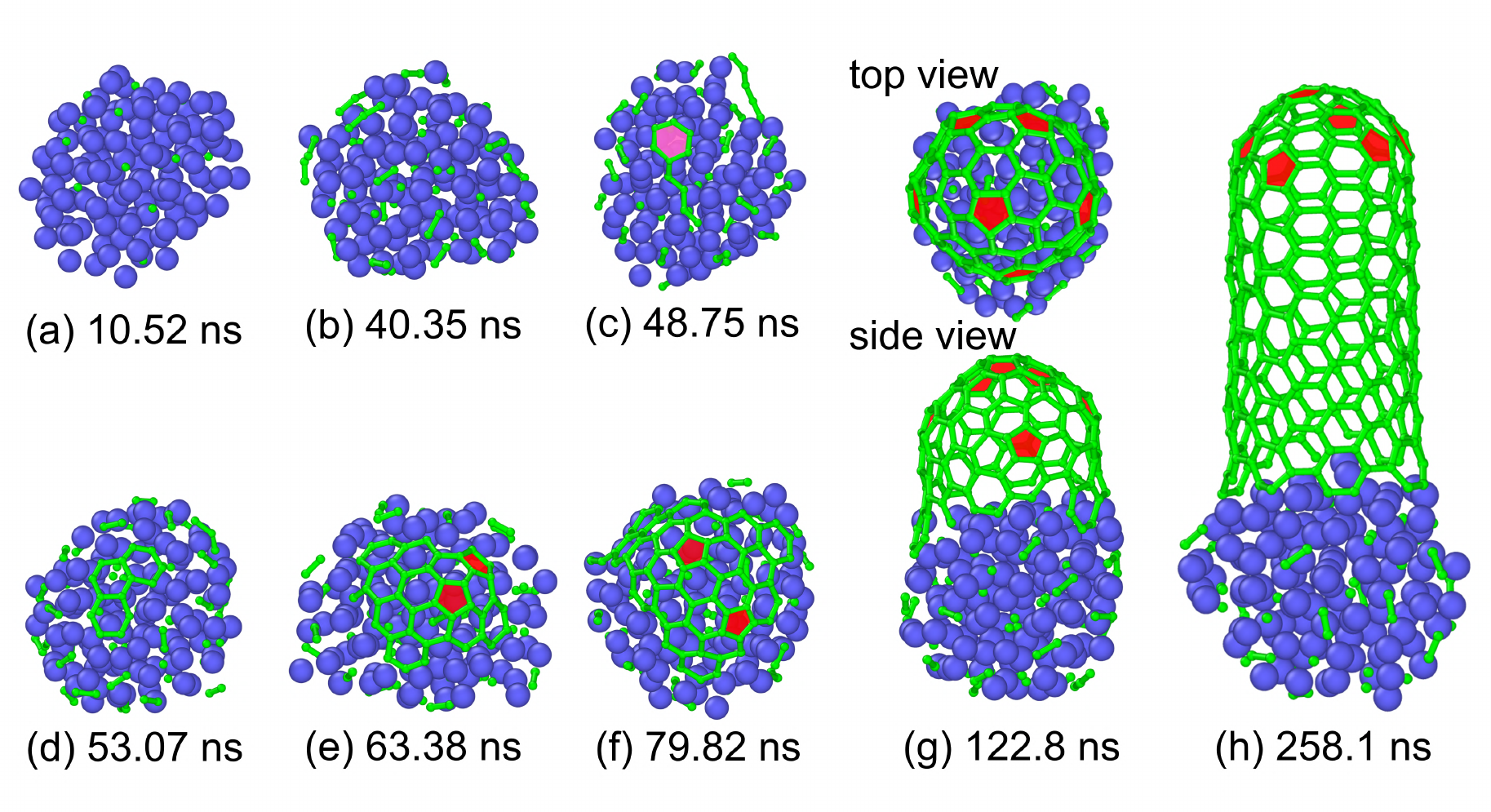}
\end{minipage}\\
\begin{minipage}[b]{0.40\columnwidth}
 \centering
\includegraphics[clip,scale=0.48]{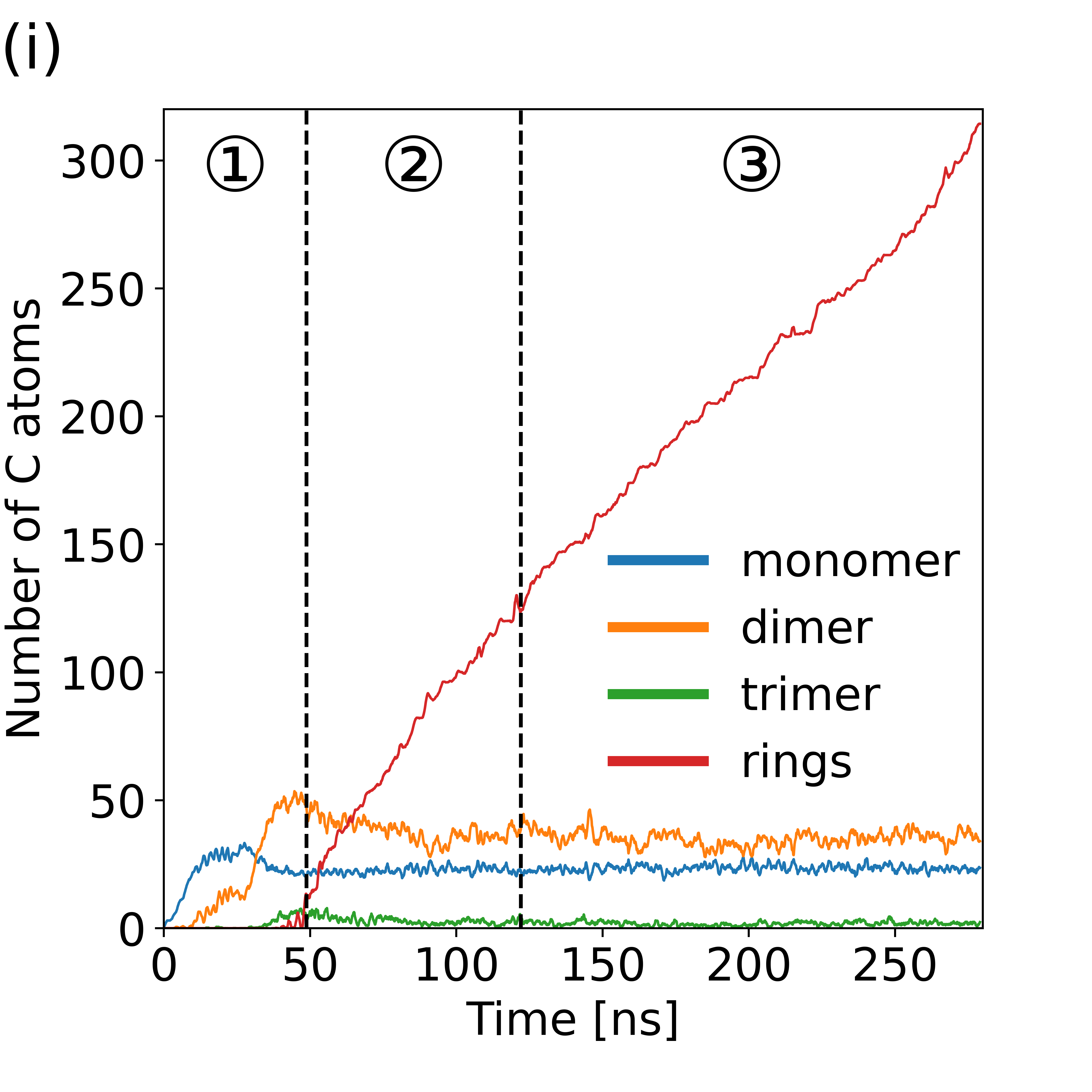}
\end{minipage}
\begin{minipage}[b]{0.40\columnwidth}
\centering
\includegraphics[clip,scale=0.48]{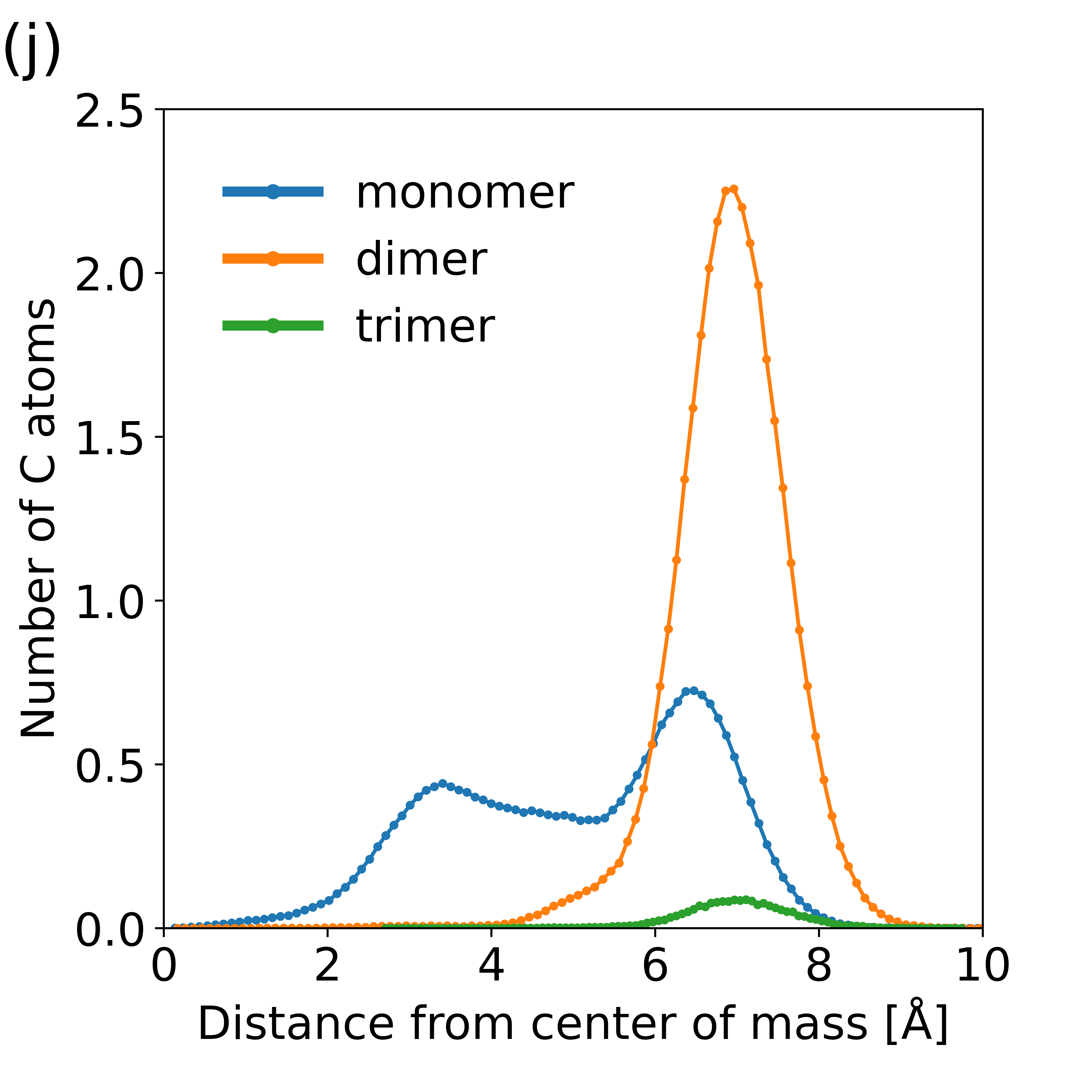}
\end{minipage}
\caption{(8,7) SWCNT growth on Fe$_{120}$ at 1500 K. Snapshots in the (a-b) saturation, (c-g) cap formation, and (h) sidewall elongation stage. The red-colored rings denote the pentagons. (i) Time evolution of the number of C atoms belonging to monomers, dimers, trimers, and rings. The circled numbers denote the growth stages. {\large \textcircled{\small 1}}, {\large \textcircled{\small 2}}, and {\large \textcircled{\small 3}} are saturation stage, cap formation stage, and sidewall elongation stage, respectively. (j) Radial distribution function of C atoms in monomers, dimers, and trimers from the center of mass of the Fe nanoparticles. The radial distribution is averaged over the sidewall elongation stage. Two C atoms within 2.0 \r{A} are regarded as being bonded.}
\label{fig:MD}
\end{figure}

Figure \ref{fig:MD}a-h shows the SWCNT growth process at 1500 K on Fe$_{120}$. At this temperature, SWCNTs grew through the typical growth stages; carbon saturation, cap formation, and side wall elongation, as reported in computational\cite{doi:10.1021/acs.jpcc.7b12687,doi:10.1021/acsnano.8b09754} and experimental\cite{doi:10.1021/nl080452q} studies.
In the saturation stage, supplied C atoms in the gas phase were adsorbed on the catalyst surface and preferentially occupied the subsurface sites (Figure \ref{fig:MD}a). When the subsurface sites were fully occupied, carbon atoms formed small molecules such as dimer and trimer on the surface (Figure \ref{fig:MD}b). The number of small molecules such as dimers and trimers continued to increase, resulting in the carbon supersaturation of the catalyst (see Figure \ref{fig:MD}i).

In the cap formation stage, the small molecules started bonding and forming short chains as the catalyst supersaturated. The cyclization of the chain led to the first ring formation (Figure \ref{fig:MD}c). Note that as a trend over the catalyst sizes from Fe$_{55}$ to Fe$_{120}$, the first ring is likely to be a hexagon, which is contrary to the previous quantum mechanical molecular dynamics (QM/MD) simulations\cite{doi:10.1021/ar100064g}. Additional polygons were formed next to the first hexagon and a small graphitic island was formed (Figure \ref{fig:MD}d). The island was enlarged as C atoms were supplied (Figure \ref{fig:MD}e,f), and the sixth pentagon formation finished the cap formation (Figure \ref{fig:MD}g). The number of these small molecules such as dimer and trimer continued to decrease as they transformed into rings, gradually resolving the supersaturation of the catalyst during this stage (see Figure \ref{fig:MD}i).

In the sidewall elongation stage, C atoms diffused on the catalyst surface and were bonded with the SWCNT edge, forming a series of hexagons. In this stage, defects were often formed on the edge, but healed into hexagons by bond recombination, resulting in the defect-free SWCNT (Figure \ref{fig:MD}h). During this stage, the number of carbon molecules in the catalyst was constant (see Figure \ref{fig:MD}i), and the monomers existed both on the surface and in the subsurface region, while larger molecules containing more than two carbon atoms appeared only on the surface region (see Figure \ref{fig:MD}j). 

\begin{figure}[H]
   \begin{minipage}[b]{0.40\columnwidth}
    \centering
    \includegraphics[clip,scale=0.48]{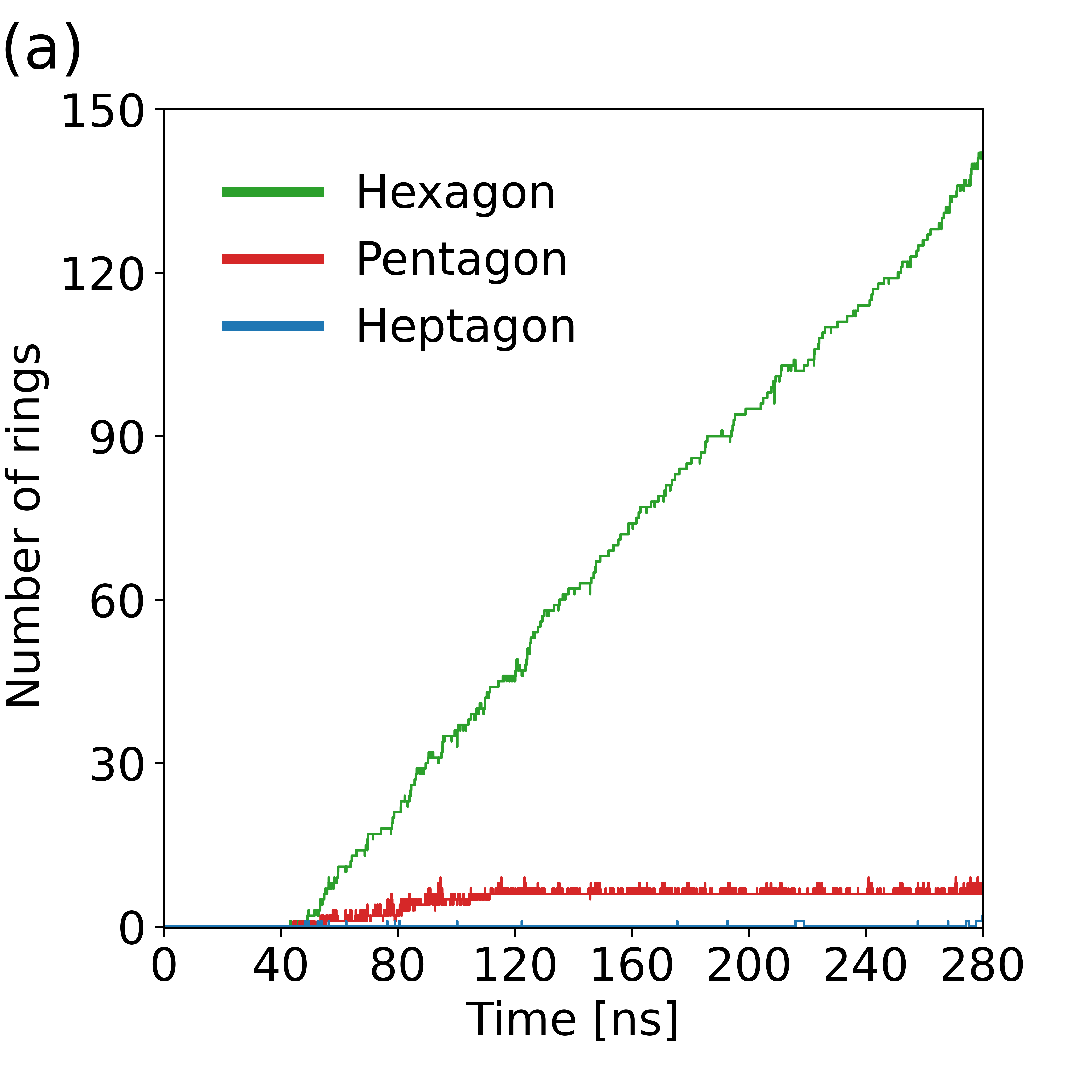}
  \end{minipage}
  \begin{minipage}[b]{0.40\columnwidth}
    \centering
    \includegraphics[clip,scale=0.48]{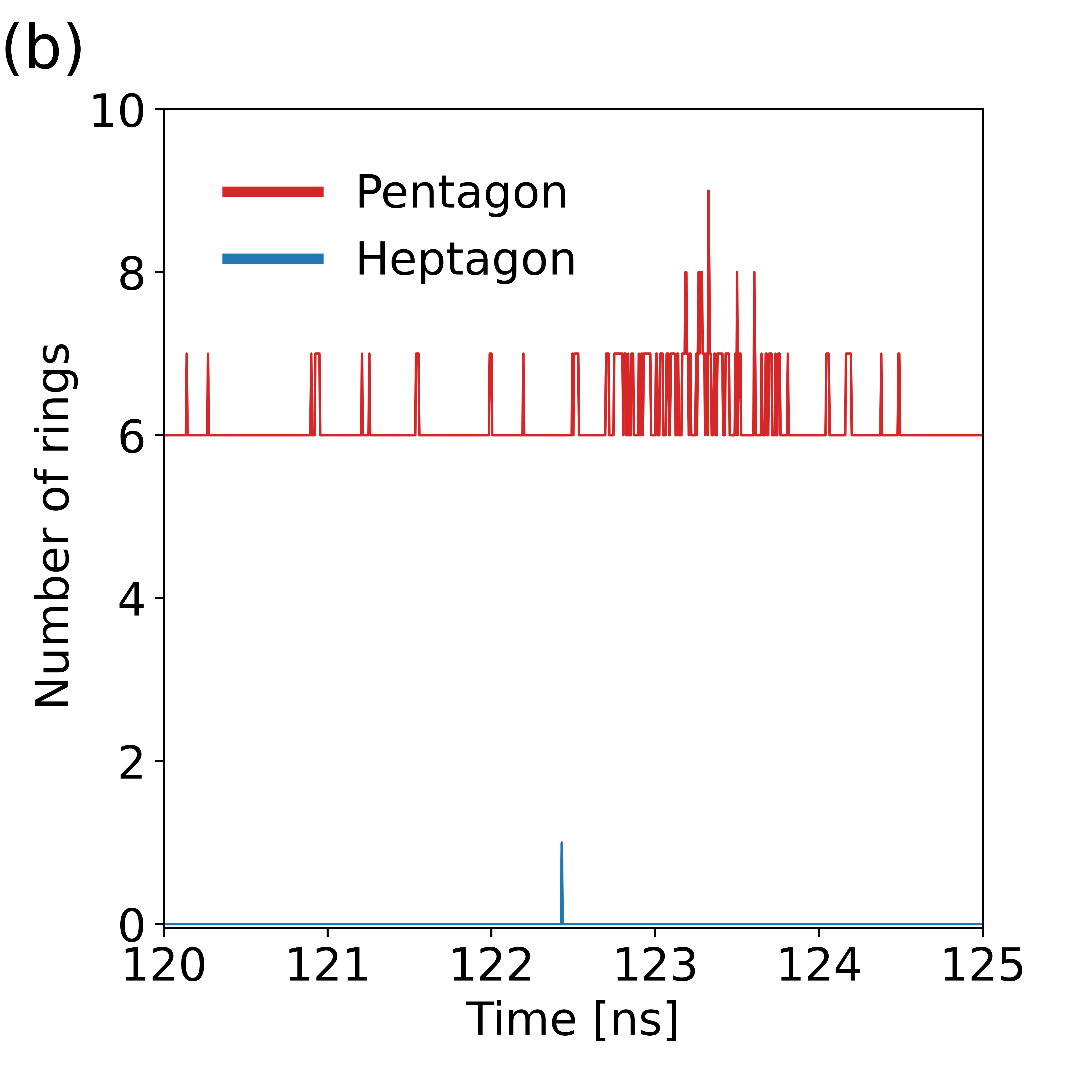}
  \end{minipage}
  \begin{minipage}[b]{1.0\columnwidth}
      \centering
      \includegraphics[clip,scale=0.55]{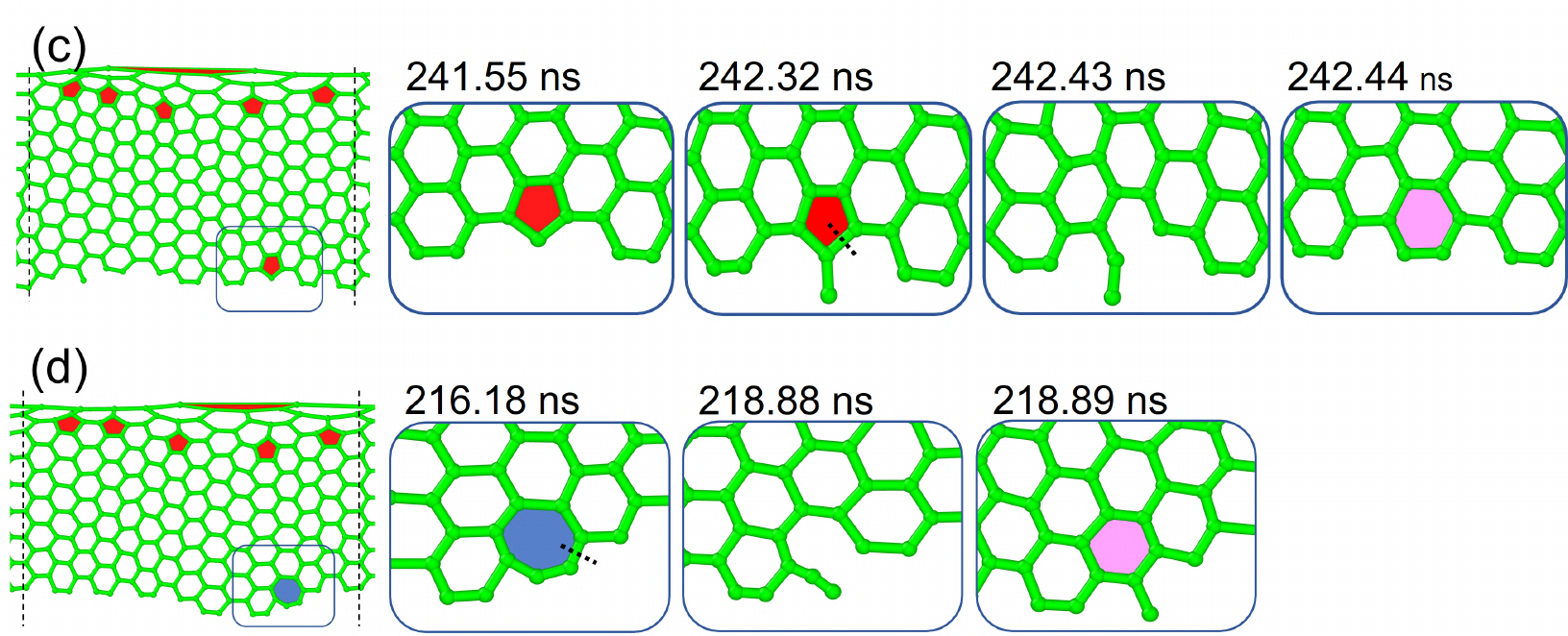}
   \end{minipage}
\caption{(a) Time evolution of the numbers of hexagons, pentagons, and heptagons. (b) An enlarged view of (a). (c) Reformation of a pentagon into a hexagon on the (8,7) SWCNT. (d) Reformation of a heptagon into a hexagon on the (8,7) SWCNT. The red and blue polygons denote the pentagon and heptagon, respectively. The dotted lines indicate the bonds that are about to break.}
\label{fig:heal}
\end{figure}

Figure \ref{fig:heal}a, and b compare the time evolution of the numbers of hexagons, pentagons, and heptagons. The number of hexagons increased at a constant rate after the first formation. Note that in the case of the (8,7) SWCNT, the hexagon formation rate was 0.61 ring/ns and the growth rate in length was $10^{4}$ $\mu$m/s, which is about $10^{2}$ times higher than the experimental growth rates of SWCNTs\cite{doi:10.1021/cm903866z} and about 3 times higher than that of the previous SWCNT growth simulation performed using the classical interatomic potential by some of us\cite{doi:10.1021/acsnano.8b09754}. The number of pentagons increased to six during the cap formation and then fluctuated but never fell below six during the sidewall elongation, indicating that the cap's pentagons were maintained. The number of heptagons was almost always zero; even when it reached one, it immediately returned to zero.
Figure \ref{fig:heal}c, and d show the detailed mechanism of defect healing on the SWCNT edge. A pentagon was formed on the edge and healed into hexagons by adding a monomer and C-C bond reformation (Figure \ref{fig:heal}c). A heptagon was formed less frequently than a pentagon, and quickly healed to a hexagon by rejecting a monomer (Figure \ref{fig:heal}d). Such defect healing close to the nanotube-catalyst interface resulted in defect-free SWCNT growth. 

\subsubsection{Statistical analysis of edge configurations}
During the growth of CNTs, the edge configuration dynamically changed depending on the local structure such as armchair and zigzag edges. According to the thermodynamic model proposed in previous studies\cite{doi:10.1126/science.aat6228,doi:10.1021/acsnano.2c07388}, the probability distribution of the edge configurations of a nanotube is governed by the interfacial Gibbs energy between the nanotube and the catalyst. 
To verify the consistency between the MD simulations and the thermodynamic model, we compared the probability distribution of appearance of edge configurations of a nanotube predicted by the thermodynamic model and frequency distribution obtained from the simulated SWCNT growth.

\begin{figure}[H]
\begin{minipage}[h]{0.49\columnwidth}
\raggedleft
\includegraphics[clip,scale=0.50]{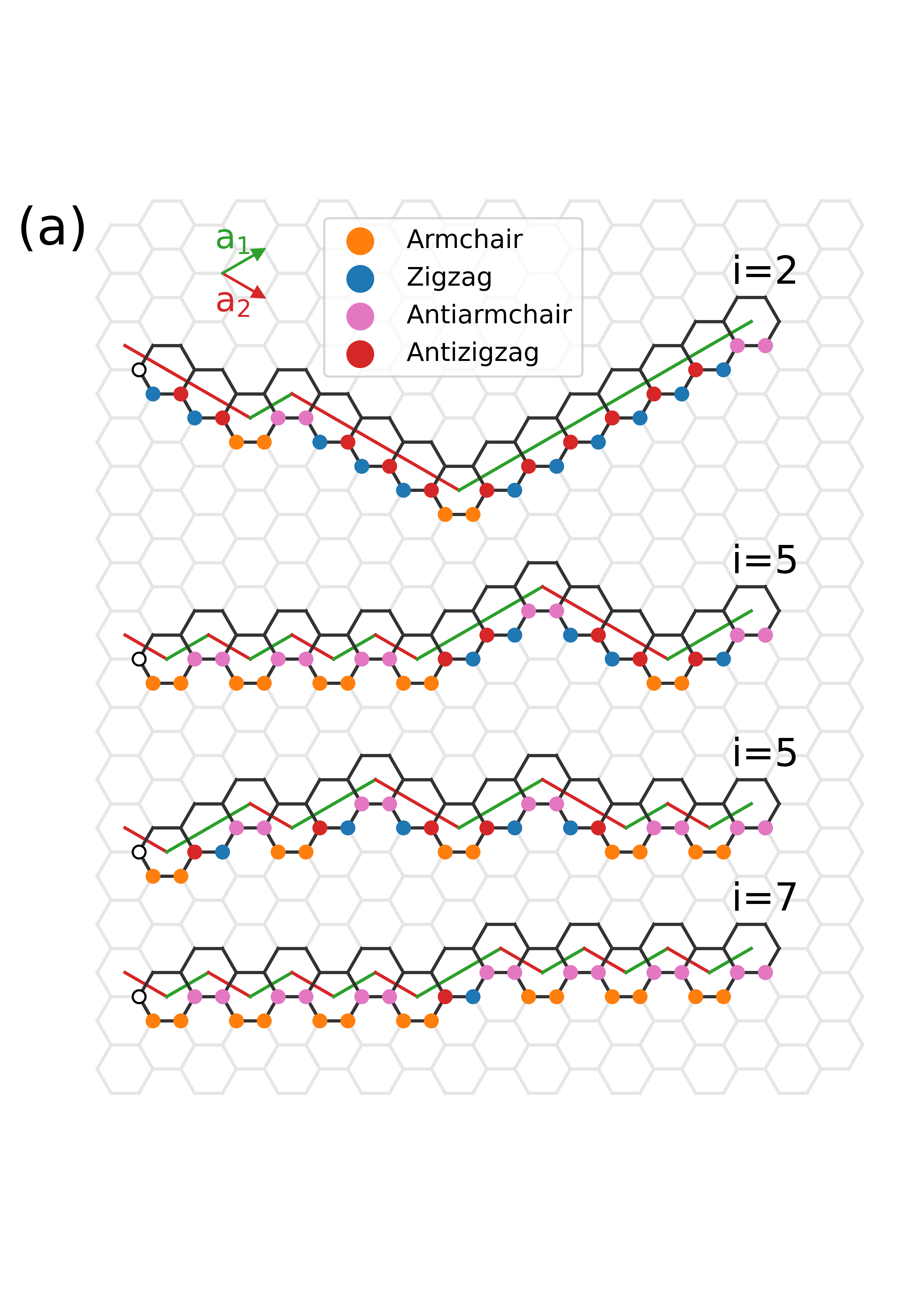}
\end{minipage}
\begin{minipage}[h]{0.49\columnwidth}
\centering
\includegraphics[clip,scale=0.15]{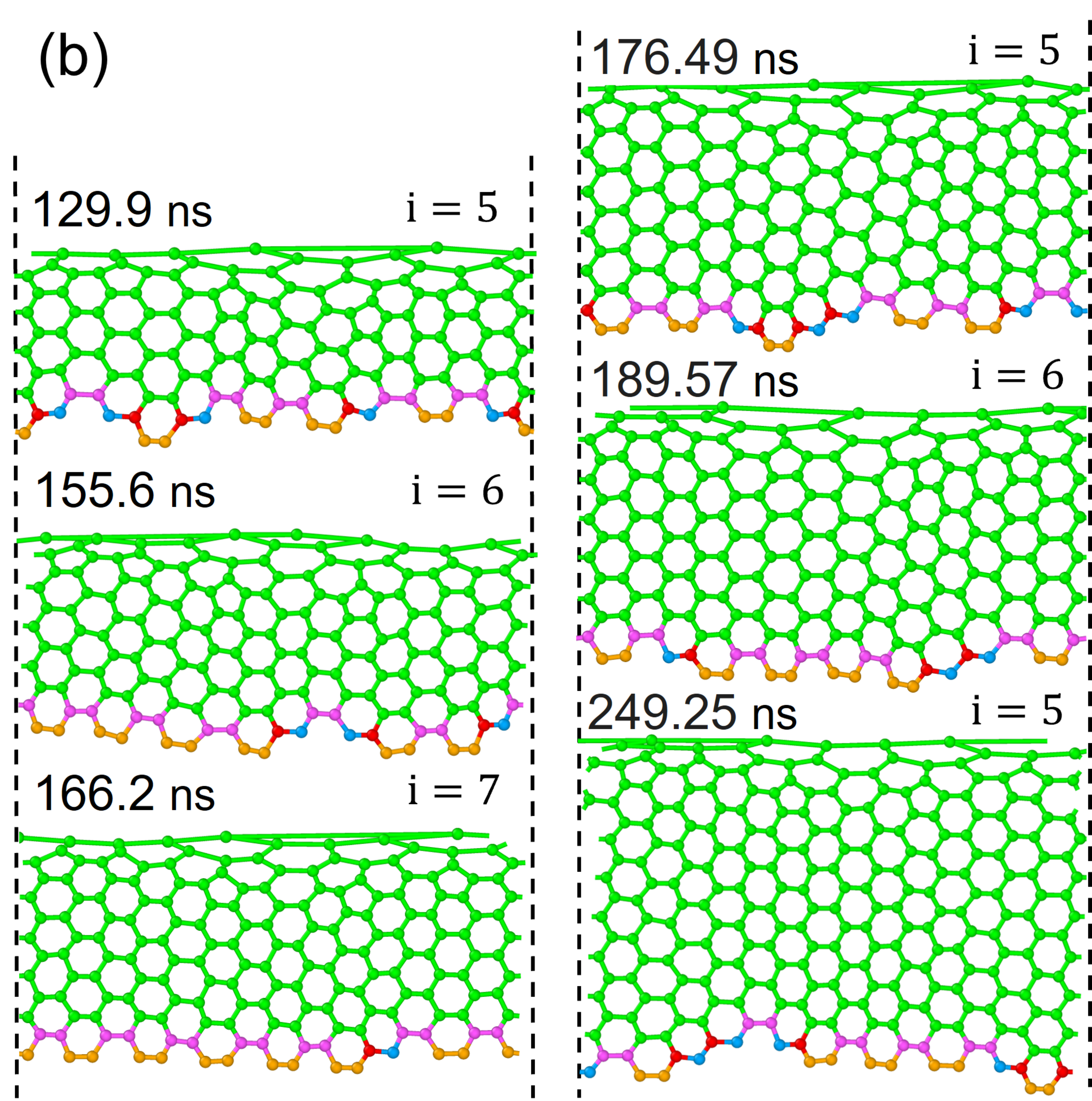}
\end{minipage}
\begin{minipage}[h]{0.49\columnwidth}
\centering
\includegraphics[clip,scale=0.48]{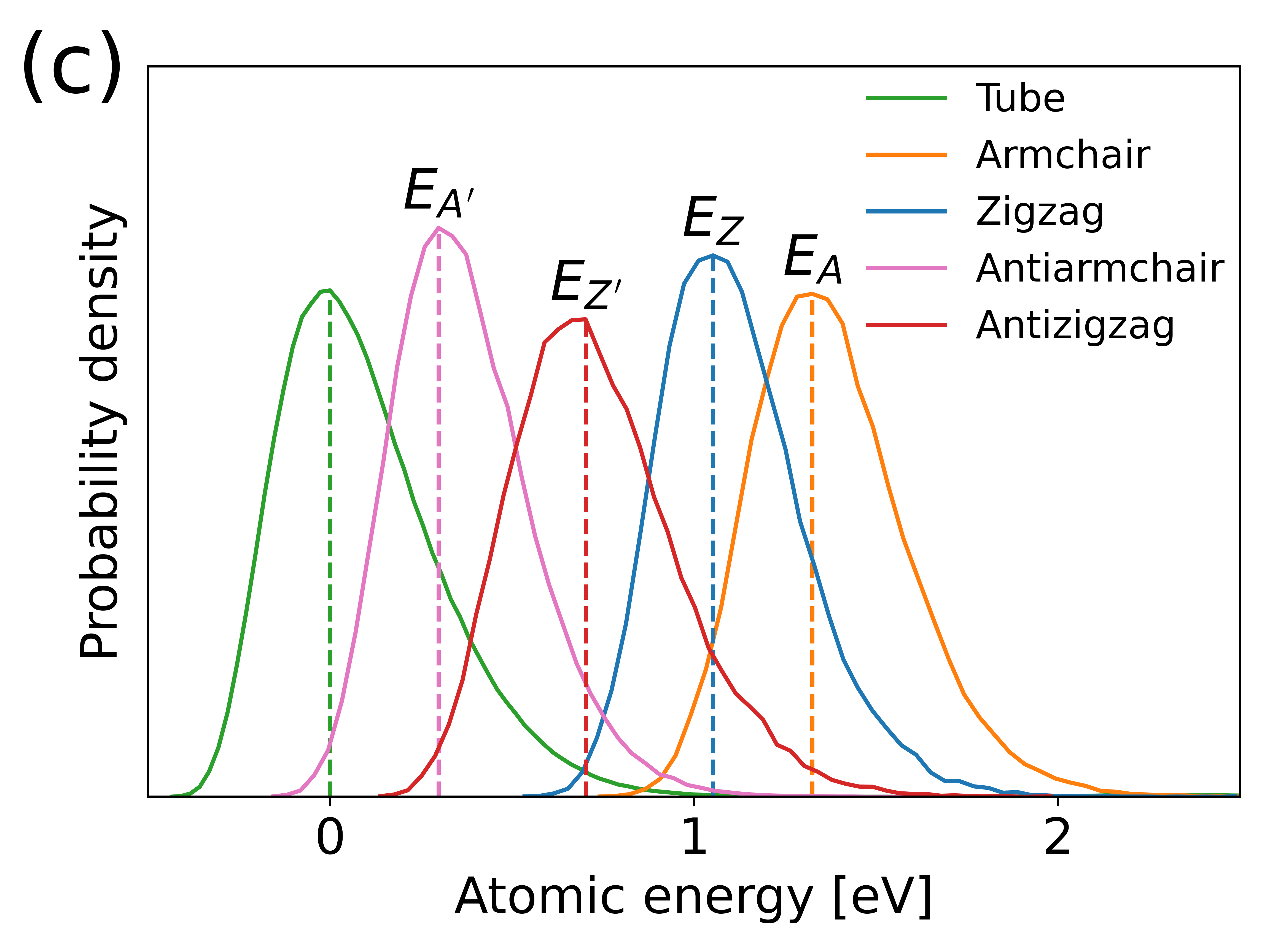}
\end{minipage}
\begin{minipage}[h]{0.49\columnwidth}
\centering
\includegraphics[clip,scale=0.48]{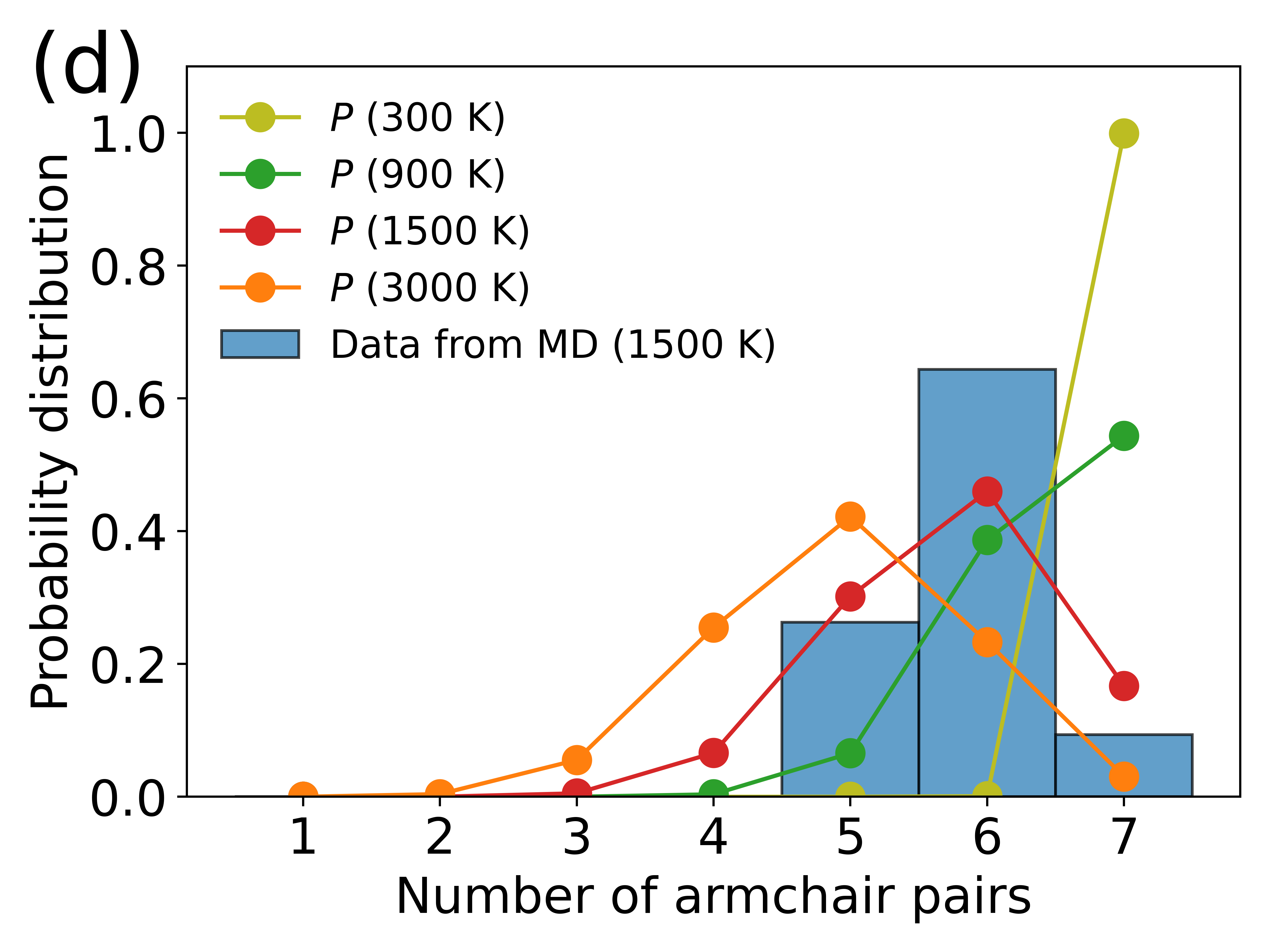}
\end{minipage}
\caption{(a) (8,7) SWCNT edges with different numbers of armchair pairs $i$ and classified edge atoms. The orange, blue, pink, and red circles denote armchair, zigzag, antiarmchair, and antizigzag atoms, respectively. (b) Developed snapshots obtained from the (8,7) SWCNT growth simulation. The orange, blue, pink, and red balls denote armchair, zigzag, antiarmchair, and antizigzag edge atoms, respectively. The green balls denote tube atoms that are not classified as edge atoms. (c) Probability density for the atomic energy of tube (green), armchair (orange), zigzag (blue), antiarmchair (pink), and antizigzag (red) atoms obtained from the MD simulation of (8,7) SWCNT growth. The atomic energy values are given relative to the peak position for the tube atoms. (d) The frequency distribution of edge configurations that appeared in the simulated (8,7) SWCNT growth at 1500 K and the probability distribution $P(n,m,i)$ at 300, 900, 1500, and 3000 K.} 
\label{fig:edgeshape}
\end{figure}

First, we classify the edge atoms in an $(n,m)$ nanotube into armchair, zigzag, antiarmchair, and antizigzag atoms, as shown in Figure \ref{fig:edgeshape}a. If the nanotube has the smallest number of edge atoms without vacancies and adatoms, $(n+m)$, and has $i$ armchair atom pairs, the number of edge atoms in each category satisfies the following equations: 
\begin{equation}
\label{eq:rule1}
N_{Z}+N_{A}+N_{Z'}+N_{A'} = 2n+2m,
\end{equation}
\begin{equation}
N_{Z} = N_{Z'} = n+m-2i,
\end{equation}
\begin{equation}
\label{eq:rule2}
N_{A} = N_{A'} = 2i,
\end{equation}
where $N_{Z}$, $N_{A}$, $N_{A'}$, and $N_{Z'}$ are the number of armchair, zigzag, antiarmchair, and antizigzag atoms, respectively. As an example, we consider the edge configurations satisfying Eq.\ref{eq:rule1}-\ref{eq:rule2} in the (8,7) SWCNT growth simulation, as shown in Figure \ref{fig:edgeshape}b. In the DeepMD architecture, the potential energy of the system is calculated as a sum of the atomic energy predicted from the local atomic environment of each atom. Thus, the interfacial energy can be estimated from the difference between the atomic energy of edge atoms and that of atoms in the nanotube. Figure \ref{fig:edgeshape}c shows the probability density for the atomic energy of armchair, zigzag, antiarmchair, antizigzag edge atoms, and tube atoms, which are 3-coordinated carbon atoms 
not categorized as edge atoms, obtained from the (8,7) SWCNT growth simulation. The density of states for each category shows a unimodal distribution, and the peak positions for the edge atoms are shifted in a positive direction compared to that for the tube atoms, indicating the energy loss due to the nanotube-catalyst interface. With ignoring the effects of the tube atoms, the interfacial energy $E_\mathrm{Int}(n,m,i)$ of an $(n,m)$ nanotube with $i$ armchair atom pairs can be approximately calculated from the difference in the peak positions of the distributions for the edge atoms as
\begin{align}
E_\mathrm{Int}(n,m,i) &=(n+m-2i)(E_{Z}+E_{Z'})+2i(E_{A}+E_{A'}) \\
&= (n+m)(E_{Z}+E_{Z'}) + 2i \Delta E_\mathrm{Int},
\end{align}
where $E_{Z}$, $E_{Z'}$, $E_{A}$, and $E_{A'}$ are the peak positions for zigzag, antizigzag, armchair, and antiarmchair atoms relative to that for tube atoms and $\Delta E_\mathrm{Int}=E_{A} + E_{A'} - E_{Z} - E_{Z'}$.  The specific value of $\Delta E_\mathrm{Int}$ depends on the condition of the nanotube-catalyst interface. In the (8,7) SWCNT growth simulation, $E_{A}$, $E_{A'}$, $E_{Z}$, and $E_{Z'}$ were 1.325, 0.298, 1.052, and 0.702 eV, respectively, and $\Delta E_\mathrm{Int}$ was thus $-0.131$ eV.
Using the interfacial energy $E_\mathrm{Int}(n,m,i)$, the canonical partition function $Z(n,m)$ for the edge configurations is calculated as
\begin{equation}
Z(n,m) = \sum_{i=1}^{m} g(n,m,i) \exp \left(-\frac{E_\mathrm{Int}(n,m,i)}{k_\mathrm{B}T} \right),
\end{equation}
where $T$ is the temperature of the system and $g(n,m,i)$ is the number of unique edges of an $(n,m)$ nanotube with $i$ armchair pairs. 
The probability distribution of appearance of edge configurations $P(n,m,i)$ is then calculated as
\begin{align}
\label{eq:dist}
P(n,m,i) &= \frac{g(n,m,i) \exp \left(-E_\mathrm{Int}(n,m,i)/k_\mathrm{B}T \right)}{Z(n,m)} \\
%&= \frac{g(n,m,i) \exp (-(n+m)(E_{Z}+E_{Z^{'}})/k_\mathrm{B}T) \exp (-2i\Delta E_\mathrm{Int}/k_\mathrm{B}T)}{\sum_{i=1}^{m} g(n,m,i) \exp (-(n+m)(E_{Z}+E_{Z^{'}})/k_\mathrm{B}T) \exp (-2i\Delta E_\mathrm{Int}/k_\mathrm{B}T)} \\
%&= \frac{g(n,m,i) \exp (-(n+m)(E_{Z}+E_{Z^{'}})/k_\mathrm{B}T) \exp (-2i\Delta E_\mathrm{Int}/k_\mathrm{B}T)}{\exp (-(n+m)(E_{Z}+E_{Z^{'}})/k_\mathrm{B}T)\sum_{i=1}^{m} g(n,m,i)  \exp (-2i\Delta E_\mathrm{Int}/k_\mathrm{B}T)} \\
&= \frac{g(n,m,i) \exp (-2i\Delta E_\mathrm{Int}/k_\mathrm{B}T)}{\sum_{i=1}^{m} g(n,m,i) \exp (-2i\Delta E_\mathrm{Int}/k_\mathrm{B}T)}.
\end{align}

The degeneracy $g(n,m,i)$, i.e. the number of edges of a $(n,m)$ nanotube with $i$ armchair pairs, depends on the number of discernible species defining the nanotube edge. In the previous models, the edge was defined as a set of zigzag atoms and armchair atoms pairs, assuming either a constant number of armchair and zigzag species \cite{doi:10.1126/science.aat6228} or letting them fluctuate to account for oblique edges with respect to the tube axis \cite{doi:10.1021/acsnano.2c07388}. In the latter, $g_1(n,m,i)$ was calculated as the number of combinations of $i$ armchair sites and $n+m-2i$ zigzag sites:
\begin{equation}
g_{1}(n,m,i) = \binom{n+m-i}{i}=\frac{(n+m-i)!}{i!(n+m-2i)!},
\end{equation}
and in the former, $g_{0}(n,m)$ was the limiting case for $i=m$. 
\begin{equation}
g_{0}(n,m) = \frac{n!}{m!(n-m)!},
\end{equation}

These simple approximations of $g(n,m,i)$ emphasized the role of the edge configurational entropy in carbon nanotube growth, but were incomplete in the sense that the role of antiarmchair and antizigzag species that connect armchair and zigzag ones was overlooked. It is however possible to include all these four species and obtain an exact evaluation of the degeneracy $g_2(n,m,i)$ by defining  edge configurations as sets of $n$ $a_{1}$ vectors and $m$ $a_{2}$ vectors, the unit vectors of a graphene unit cell.
The derivation of $g_{2}(n,m,i)$ is explained in Supporting Information and Figures S6 and S7.
Another mathematically rigorous proof was provided by F. Ducastelle \cite{Ducastelle}.
\begin{equation}
\label{eq:g2}
g_{2}(n,m,i) = \binom{m}{i}\binom{n-1}{i-1}+\binom{n}{i}\binom{m-1}{i-1}= \frac{(n+m)i}{nm}\frac{n!m!}{(n-i)!(m-i)!(i!)^{2}},
\end{equation}
which satisfies
\begin{equation}
\sum_{i}^{m}g_{2}(n,m,i) = \binom{n+m}{n}=\frac{(n+m)!}{n!m!}.
\end{equation}

Figure \ref{fig:edgeshape}d shows the probability distribution $P(n,m,i)$ with $\Delta E_\mathrm{Int}=-0.131$ eV and $T=$ 300, 900, 1500, and 3000 K. The frequency distribution obtained from the simulated (8,7) SWCNT growth at 1500 K is also shown for comparison. The predicted distribution at 1500 K is consistent with the frequency distribution from the growth simulation at 1500 K in terms of the order of probability of appearance. 
%As the temperature decreases, the probability of having more armchair pairs increases, so that in the limit $T\to0$ the edges are restricted to only those with $i=m$, resulting in the spiral growth of SWCNTs, where only armchair-zigzag kinks serve as active sites for growth. 
The same analysis was performed on the (9,9) SWCNT, as shown in Figure S8. For the (9,9) SWCNT, $\Delta E_\mathrm{Int}$ was calculated as $-0.083$ eV. As in case of the (8,7) SWCNT, the predicted distribution $P(n,m,i)$ at 1500 K is also consistent with the frequency distribution obtained from the (9,9) SWCNT growth simulation.
The temperature dependence and the interfacial energy dependence of $P(n,m,i)$ calculated with $g_{1}$ and $g_{2}$ for various chiralities are shown in Figure S9 and Figure S10, respectively. The difference in counting edges leads to qualitative differences in the temperature and interfacial energy dependence of $P(n,m,i)$. As a general trend, as the temperature decreases, the distribution of edge configurations becomes narrower and ultimately only configurations with $m$ armchair pairs survive, as expected when the configurational entropy goes to zero. 

Note that on the assumption that nanotubes grow with the edge configurations satisfying Eq.\ref{eq:rule1}-\ref{eq:rule2}, $(n,1)$ nanotubes should grow by the kink-running process regardless of the values of $T$ and $\Delta E$ as reported in the previous MD simulations by Yoshikawa et. al.\cite{doi:10.1021/acsnano.8b09754} because only edge configurations with one antiarmchair site are allowed, as shown in Figures S9 and S10.

\subsubsection{Effect of ring formation at antizigzag sites}
\begin{figure}[H]
\begin{minipage}[b]{1.0\columnwidth}
\centering
\includegraphics[clip,scale=0.40]{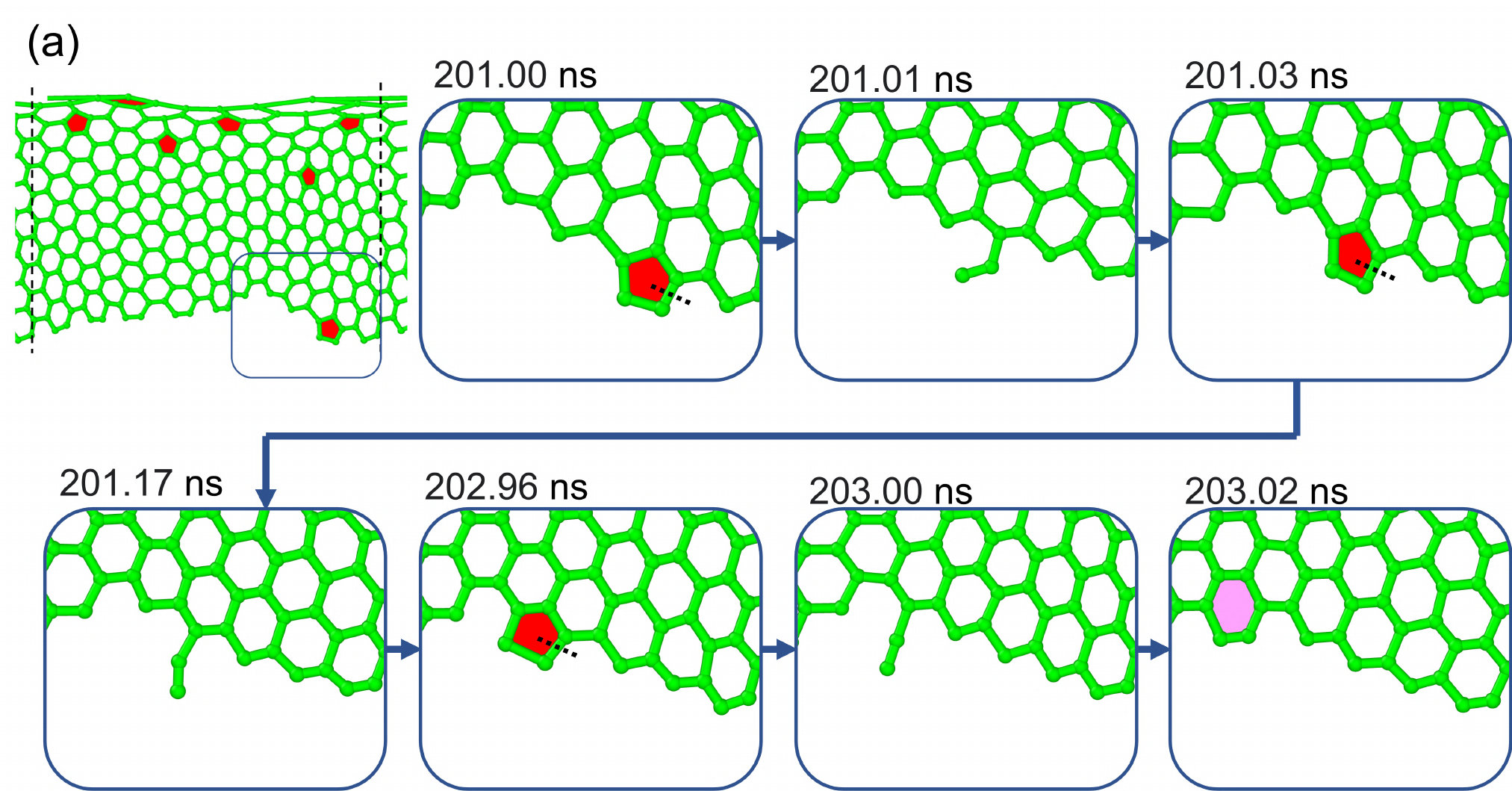}
\end{minipage}
\begin{minipage}[b]{1.0\columnwidth}
\centering
\includegraphics[clip,scale=0.40]{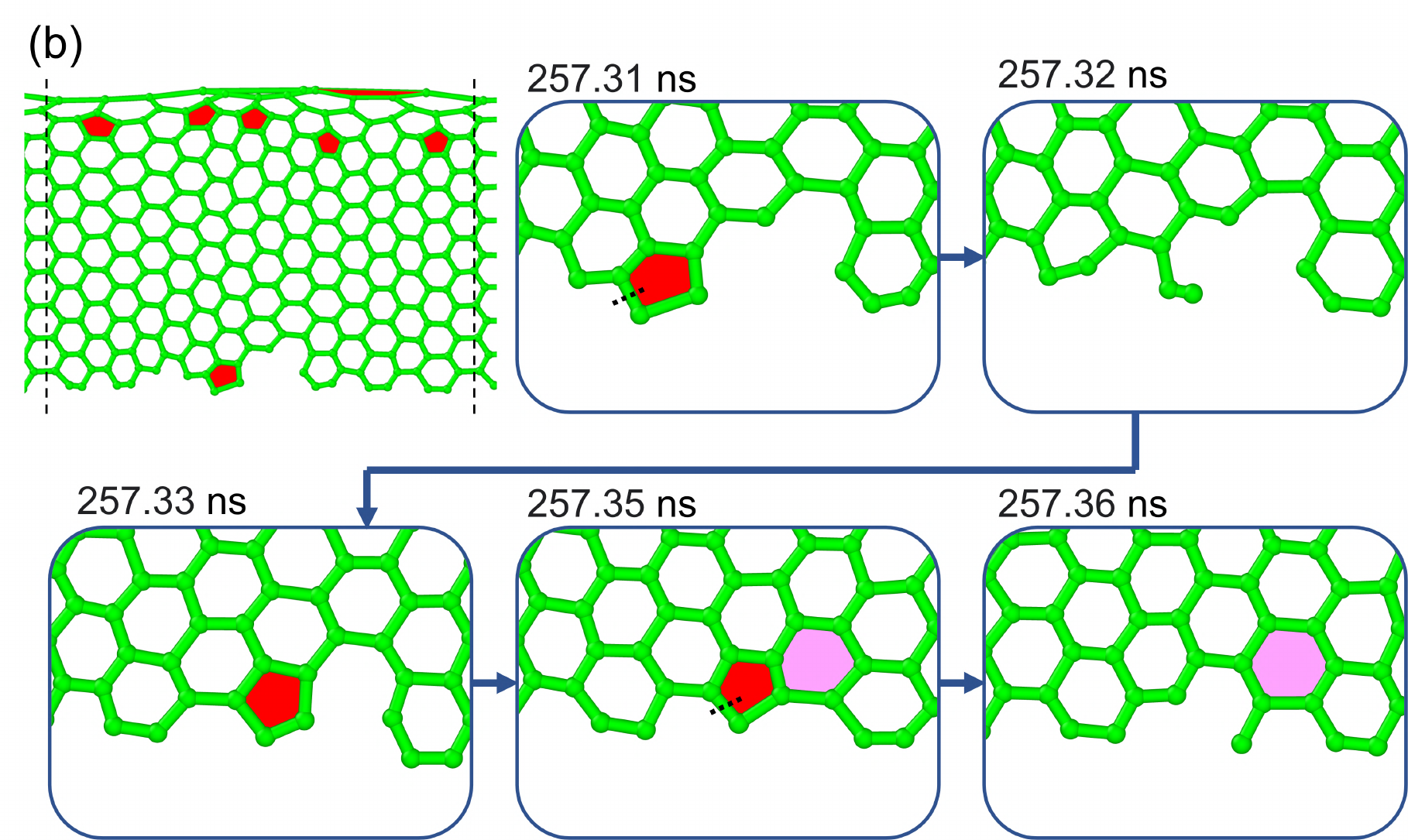}
\end{minipage}
\caption{(a) Adatom diffusion along the zigzag edge of the (10,7) SWCNT. (b) Edge vacancy healing by adatom diffusion on the (8,7) SWCNT. The dotted lines indicate the bonds that are about to break.} 
\label{fig:diffuse}
\end{figure}

The above discussions have been based on the assumption that rings are formed only at antiarmchair sites, but in the MD simulations there are cases where rings are formed at antizigzag sites as well, which is also important for the growth mechanism of SWCNTs.
At antizigzag sites, adatoms often formed a protruding pentagon which is thermodynamically unfavorable. 
The adatoms frequently jumped to adjacent antizigzag sites with repeated bonding and dissociation like a trapeze. When they reached an antiarmchair site, they formed a hexagon and are stabilized (Figure \ref{fig:diffuse}a). Such adatom diffusion healed the vacancies, preventing defect formation (Figure \ref{fig:diffuse}b).

\begin{figure}[H]
\begin{minipage}[b]{1.0\columnwidth}
\centering
\includegraphics[clip,scale=0.36]{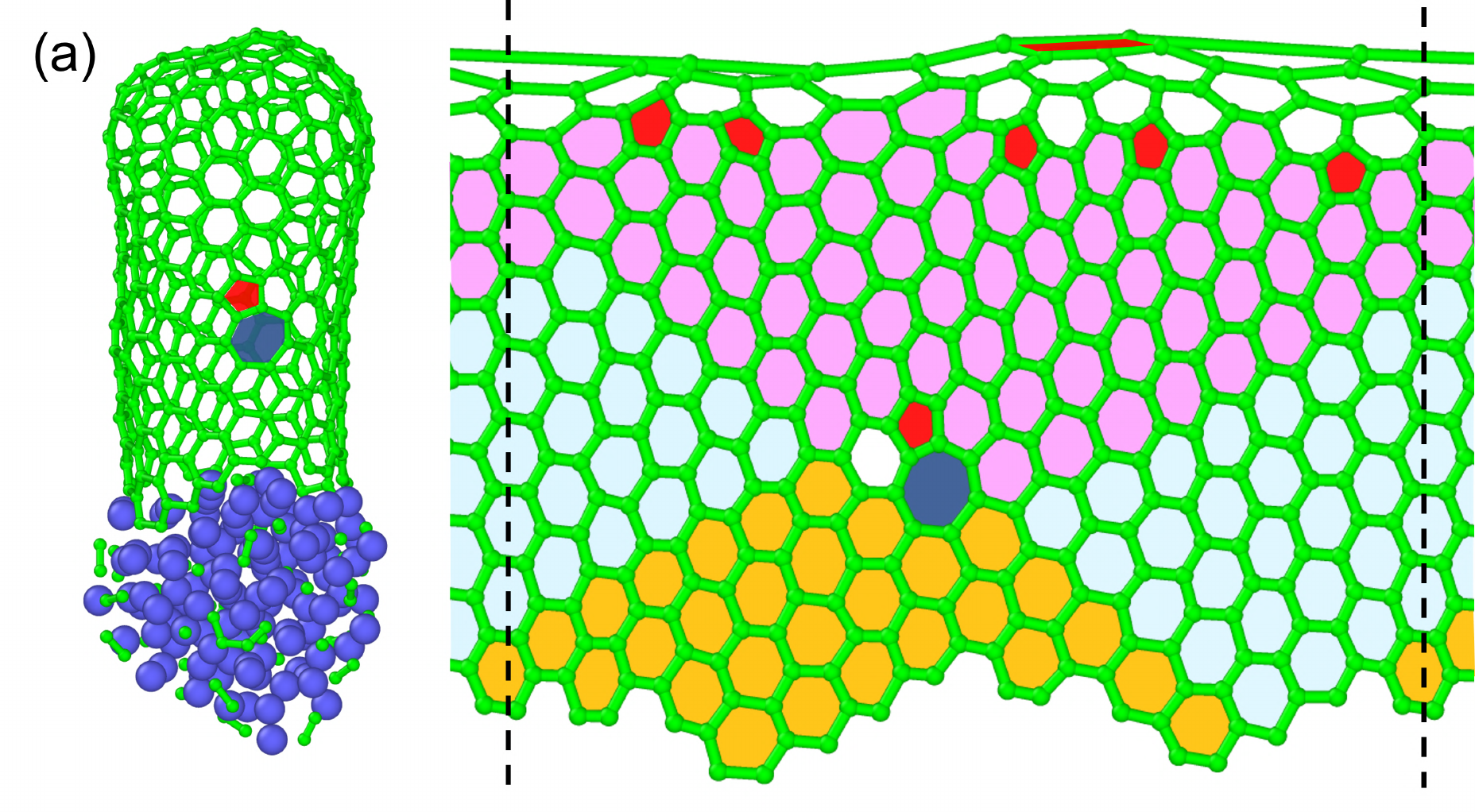}
\end{minipage}
\begin{minipage}[b]{1.0\columnwidth}
\centering
\includegraphics[clip,scale=0.44]{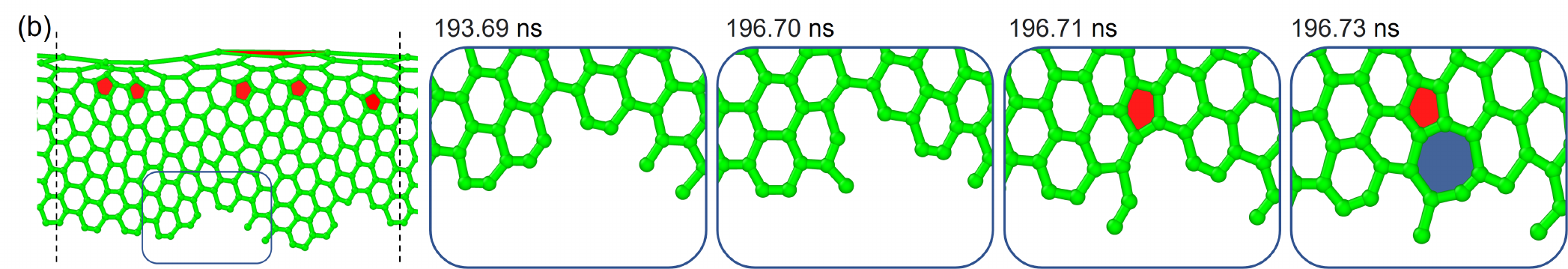}
\end{minipage}
\caption{An SWCNT with a 5-7 defect-mediated chirality change at 1500 K on Fe$_{120}$. (a) Snapshot and developed view of the SWCNT with chirality change from (11,7) to (10,7) after 280 ns of carbon supply. (b) 5-7 defect formation process. The pink and orange areas are (11,7) and (10,7) sidewalls, respectively. The light blue areas can be defined as (11,7) and (10,7) since there are two possible translation vectors. The chirality of the white areas can not be assigned. The red and blue polygons denote the pentagons and heptagons, respectively. } 
\label{fig:disloc}
\end{figure}

In the MD simulations, unhealed vacancies occasionally resulted in defects as shown in Figure \ref{fig:disloc}, because the growth rate in the simulations was much higher than that in the experiments and there is not enough time for vacancy to heal. In the experiments, however, SWCNTs are considered to grow while maintaining smooth edges through adatom diffusion, resulting in a low probability of defect formation.

On armchair edges, pentagons were quickly healed to hexagons by adding monomers because such pentagons had dangling bonds and were highly reactive, as shown in Figure \ref{fig:heal}a. However, on zigzag edges, pentagons without dangling bonds were often formed at edge vacancies (Figure \ref{fig:disloc}b). Such pentagons were less reactive, resulting in less susceptibility to monomer addition and more chances of being left unhealed inside the sidewalls. Heptagons were often formed below the pentagons to compensate for the curvature change by the pentagon formation, resulting in pairs of adjacent pentagons and heptagons (5-7 defects). 5-7 defects caused chirality changes as reported in previous experimental studies\cite{doi:10.1126/science.291.5501.97,Yao2007,doi:10.1021/acsnano.8b01630}. An SWCNT with 5-7 defect-mediated chirality change is shown in Figure \ref{fig:disloc}a. The SWCNT grew to a certain length with the initial chirality (11,7) determined by the cap and changed to (10,7) by a 5-7 defect. We also observed the chirality change from (10,3) to (8,3) by two 5-7 defects (Figure S11).

\section{Conclusions}
\noindent
\hspace{\parindent}
We performed MD simulations of defect-free SWCNT growth using the NNP. The MD simulations showed the structural evolution of nanotube edges with dynamic rearrangement of edge configurations. We compared the probability of appearance of the edge configurations with the thermodynamic model reported in previous work and demonstrated that the probability can be predicted from the nanotube-catalyst interfacial energy. It was also observed that the vacancy, which is caused by the ring formation at the anti-zigzag sites, leads to the formation of defects and that the adatom diffusion along the edges contributes to the healing of vacancies. These observations explain how a nanotube edge is formed and changed during the growth and provide a clue for realizing the chirality-controlled synthesis of SWCNTs.

\section{Method}

\subsection*{Development of the NNP}
\hspace{\parindent}
%The growth process of SWCNTs consists of the dynamic formation of carbon rings on the catalyst surface by the recombination of inter-carbon bonds. In order to reproduce the growth of SWCNTs by MD simulations, the interatomic potential should predict the energy of various morphologies of carbon including unstable structures. 
To collect dataset structures covering a wide range of PESs, we sampled three types of atomic structures: disordered structures, defective graphene, and structures obtained from SWCNT growth simulations.
The disordered structures were obtained by randomly placing atoms in a computational cell of 8-20 \r{A} per side, heating them to 10000 K by classical MD simulation using a Tersoff-type potential\cite{doi:10.1021/acs.jpcc.7b12687}, and then quenching them to 1000-5000 K (Figure S1a).
The defective graphene structures were collected by preparing graphene in a hexagonal supercell of 17.24$\times$17.24$\times$10.0 \r{A}, randomly displacing carbon atoms by 0-4 \r{A}, and relaxing them using Tersoff-type potential (Figure S1b).
The CNT growth structures were sampled from MD simulations of SWCNT growth on Fe$_{55}$ using Tersoff-type potential and NNP under development (Figure S1c). The cell size of the CNT growth structure was set to 15.0$\times$15.0$\times$15.0 \r{A} so that the length of the vacuum space is about 10 \r{A}. The numbers of atomic configurations included in the disordered, defective graphene and CNT growth datasets are 21744, 16871, and 5776, respectively. Each dataset was randomly separated into three subsets, with 10\% as the validation set, 10\% as the testing set, and the remainder as the training set. 

We performed spin-polarized DFT calculations using the projector-augmented wave (PAW) method\cite{PhysRevB.50.17953}, Perdew-Burke-Ernzerhof (PBE) generalized gradient approximation for exchange-correlation functional\cite{PhysRevLett.77.3865,PhysRevLett.78.1396}, with 400 eV plane wave energy cutoff, and the k-point mesh for the Monkhorst-Pack grid Brillouin zone sampling (N$_\mathrm{k_1}$, N$_\mathrm{k_2}$, N$_\mathrm{k_3}$) = $(3,3,3)$ for the disordered structures and $(1,1,1)$ for the defective graphene and CNT growth structures, as implemented in Vienna ab initio simulation package (VASP 5.4.4) \cite{PhysRevB.47.558,Kresse_1994,PhysRevB.54.11169,PhysRevB.59.1758}. The convergence criterion for the electronic self-consistent loop was set to 10$^{-4}$ eV.

Deep potential\cite{PhysRevLett.120.143001} was used for training the NNP. We used the "se\_e2\_a" descriptor implemented in DeePMD-kit\cite{WANG2018178}. The cutoff radius is set to $4.0$ \r{A}. The smoothing function was set between 3.5-4.0 \r{A}. For the embedding net, the sizes of the hidden layers from the input end to the output end are 25, 50, and 100, respectively. For the fitting net, three hidden layers with 240 neurons were used. Parameter optimization was performed in 2,000,000 steps. An exponentially decaying learning rate from 0.001 to 3.51$\times10^{-8}$ was applied. Comparisons of the energies and the forces between DFT and NNP are shown in Figure S2. The MAEs of the energies are 16.9 meV/atom and 16.1 meV/atom for the training set and testing set, and the MAEs of the forces are 0.268 eV/\r{A} and 0.248 eV/\r{A} for the training set and testing set, respectively.

\subsection*{SWCNT Growth Simulation}
MD simulations were performed using LAMMPS\cite{LAMMPS}. The time step for numerical integration was 0.5 fs. The temperature of Fe nanoparticles was controlled at $T$ using the Nose-Hoover thermostat\cite{doi:10.1063/1.447334,PhysRevA.31.1695}. 
The simulation box was a cubic cell with a size of 200$\times$200$\times$200 \r{A}. The periodic boundary condition was applied for each direction of the axis. At the beginning of the simulation, we placed Fe$_{55-120}$ nanoparticles at the center of the cell. After relaxation at constant temperature for 2 ns, we supplied C atoms at random positions in the cell controlling the density of C atoms in the gas phase at 8 atoms/cell. The initial velocity vectors of C atoms $\bm{v}$ were randomly determined, assuming that the temperature of the system is $T$ and the magnitude of the initial vector $|\bm{v}|$ satisfies $|\bm{v}|=\sqrt{3k_{b}T/m}$. 
In order to describe the energy barrier of the carbon feedstock decomposition, a Lennard-Jones potential was applied between non-covalent C atoms, as in previous SWCNT growth simulations\cite{doi:10.1021/acs.jpcc.7b12687,doi:10.1021/acsnano.8b09754}. OVITO\cite{Stukowski_2010} was used for the visualization of the atomic structures and the common neighbor analysis.

The developed views were made by expressing the coordinates of the SWCNT by the cylindrical coordinates $(r, \theta, \eta)$ and projecting them onto the x-z plane using the equation $x=C\times\theta$, $y=0$, $\eta=z$, where C is a constant that represents the average radius in the tube direction of the SWCNTs to be developed. The chirality $(n,m)$ of each SWCNT was determined by the positive combination of the basis vectors of the developed SWCNT, $a_{1}$ and $a_{2}$, according to the original definition of the chirality. 

%In this study, two C atoms were defined as being bonded if they were within 2.0 \r{A} of each other, and C atoms with one, two, and three C-C bonds were classified as $C_{b1}$, $C_{b2}$, and $C_{b3}$, respectively. A zigzag atom was defined as a $C_{b2}$ atom bonded to two $C_{b3}$ atoms. An armchair atom was defined as a $C_{b2}$ atom bonded to one $C_{b3}$ and one $C_{b2}$ atom. A terminating C atom was defined as a $C_{b1}$ atom bonded to a $C_{b2}$ or a $C_{b3}$ atom. When classifying C atoms, C-C bonds with terminating C atoms were ignored.

%%%%%%%%%%%%%%%%%%%%%%%%%%%%%%%%%%%%%%%%%%%%%%%%%%%%%%%%%%%%%%%%%%%%%
%% The "Acknowledgement" section can be given in all manuscript
%% classes.  This should be given within the "acknowledgement"
%% environment, which will make the correct section or running title.
%%%%%%%%%%%%%%%%%%%%%%%%%%%%%%%%%%%%%%%%%%%%%%%%%%%%%%%%%%%%%%%%%%%%%
\section*{Author information}
\subsection*{Corresponding Author}
Shigeo Maruyama - Department of Mechanical Engineering, The University of Tokyo, 7-3-1 Hongo, Bunkyo-ku, Tokyo, Japan; School of Mechanical Engineering, Zhejiang University, Hangzhou 310027, People’s Republic of China; orcid.org/0000-0003-3694-3070; E-mail: maruyama@photon.t.u-tokyo.ac.jp
\subsection*{Authors}

Ikuma Kohata - Department of Mechanical Engineering, The University of Tokyo, 7-3-1 Hongo, Bunkyo-ku, Tokyo, Japan
%; orcid.org/0000-0002-7468-8894

\noindent
Ryo Yoshikawa - Department of Mechanical Engineering, The University of Tokyo, 7-3-1 Hongo, Bunkyo-ku, Tokyo, Japan
%; orcid.org/0000-0002-4077-2174

\noindent
Kaoru Hisama - Research Initiative for Supra-Materials, Shinshu University, 4-17-1 Wakasato, Nagano, Nagano 380-8553, Japan
%; /orcid.org/0000-0002-0508-6427\\

\noindent
Christophe Bichara - Aix Marseille Univ, CNRS, Centre Interdisciplinaire de Nanoscience de Marseille, Marseille, France

\noindent
Keigo Otsuka - Department of Mechanical Engineering, The University of Tokyo, 7-3-1 Hongo, Bunkyo-ku, Tokyo, Japan

\subsection*{Author Contribution}
I. K. and S. M. conceived the project. I.K. developed the NNP, carried out the MD simulations. Data was analyzed and discussed by all authors. I. K. wrote the original manuscript, and all authors reviewed and edited it.
\subsection*{Notes}
The authors declare no competing financial interest.

\begin{acknowledgement}
We sincerely thank Dr. François Ducastelle, researcher at ONERA, France, who passed away on July 2, 2021 for his contribution to this work. He offered the mathematically rigorous derivation of the unique edge configuration in Eq. \ref{eq:g2}.
We thank Dr. Daniel Hedman for helpful discussion. 
We acknowledge Center for Computational Materials Science, Institute for Materials Research, Tohoku University for the use of MASAMUNE-IMR. Additional computational resources were provided by Research Center for Computational Science, Okazaki, Japan (Project: 24-IMS-C245). A part of this work was supported by JSPS (KAKENHI JP21KK0087, JP22H01411, JP23H00174, JP23H05443) and JST (CREST JPMJCR20B5). 

\end{acknowledgement}

%%%%%%%%%%%%%%%%%%%%%%%%%%%%%%%%%%%%%%%%%%%%%%%%%%%%%%%%%%%%%%%%%%%%%
%% The same is true for Supporting Information, which should use the
%% suppinfo environment.
%%%%%%%%%%%%%%%%%%%%%%%%%%%%%%%%%%%%%%%%%%%%%%%%%%%%%%%%%%%%%%%%%%%%%
\begin{suppinfo}

The Supporting Information is available free of charge on the ACS Publications website. Additional figures and data contain example structures of training dataset; comparison of total energies and forces between the NNP and DFT; details about the measurement of thermodynamic properties of bulk Fe; the Lindemann index of Fe nanoparticles as a function of temperature; growth trend of SWCNTs by temperature and catalyst size; detailed processes of hexagon and pentagon formation of the SWCNTs; snapshots and developed view of the grown defect-free SWCNTs; and snapshots and developed view of an SWCNT with chirality change from (10,3) to (8,3).

\end{suppinfo}

%%%%%%%%%%%%%%%%%%%%%%%%%%%%%%%%%%%%%%%%%%%%%%%%%%%%%%%%%%%%%%%%%%%%%
%% The appropriate \bibliography command should be placed here.
%% Notice that the class file automatically sets \bibliographystyle
%% and also names the section correctly.
%%%%%%%%%%%%%%%%%%%%%%%%%%%%%%%%%%%%%%%%%%%%%%%%%%%%%%%%%%%%%%%%%%%%%
\bibliography{achemso-demo}

\end{document}